\documentclass[10pt,twocolumn,twoside]{IEEEtran}
\usepackage{amsfonts,dsfont,amssymb,amsbsy,amsmath,paralist,theorem,bm,ifthen,color}
\usepackage[pdfstartview=FitH,bookmarksnumbered,unicode,bookmarksopen=true]{hyperref}
\usepackage{graphicx}
\usepackage[table]{xcolor}
\usepackage{epstopdf}
\usepackage{relsize}
\usepackage{mathrsfs}
\usepackage{bm}
\usepackage{stfloats}
\usepackage{url}
\usepackage[noadjust]{cite}
\usepackage{cases,booktabs}
\usepackage{graphicx,algorithmic,algorithm,relsize}
\usepackage[justification=centering]{caption} 

\usepackage[table]{xcolor}
\usepackage{array,tabularx, multirow, booktabs, hyperref}
\usepackage{threeparttable}
\usepackage{tablefootnote}
\usepackage[most]{tcolorbox}
\usepackage{bbm}
\usepackage{makecell} 
\usepackage{tikz}

\newcommand{\fullcircle}{%
  \tikz[baseline=-0.55ex]
  \fill (0,0) circle[radius=0.68ex];%
}

\newcommand{\partialcircle}{%
  \tikz[baseline=-0.55ex]{%
    \draw[line width=0.35pt] (0,0) circle[radius=0.68ex];
    \fill (0,0) circle[radius=0.15ex];
  }%
}

\newcommand{\emptycircle}{%
  \tikz[baseline=-0.55ex]
  \draw[line width=0.35pt] (0,0) circle[radius=0.68ex];%
}

\newcommand\st    {\ensuremath{{\rm s.t.}}}

\graphicspath{{fig/}}

\definecolor{sectbg}{RGB}{235,240,255} 
\definecolor{green}{RGB}{34	195	46}
\definecolor{red}{RGB}{220 0 0}

\usepackage{graphicx} 

\title{Pinching Antennas for Next-Generation Wireless Communications: A Network Perspective}

\author{Yanqing Xu, Shan Shan, Yongxu Zhu, Zhiguo Ding, Mugen Peng, and Xiaohu You
\\~\\

}

\date{\today}

\begin{document}

\maketitle

\begin{abstract}
Pinching antennas enable radiation points to be created and
reconfigured along extended waveguides, making the physical locations
of wireless transmission and reception controllable after
infrastructure deployment. This capability introduces a spatial degree
of freedom beyond conventional beamforming and resource allocation,
with implications that extend from individual links to network-wide
operation. This tutorial develops a network-oriented perspective on
pinching-antenna systems. We first review their system architecture,
channel characteristics, and fundamental design opportunities. We
then examine how radiation-location reconfigurability can support
environment-division multiple access by strengthening desired links
and suppressing cross-links, and discuss its roles in multi-cell
transmission, traffic offloading, and inter-cell cooperation. Moving
from instantaneous users to entire service regions, we introduce
traffic-aware and geometry-aware designs under a cell-free-inspired
multi-waveguide architecture. Representative system models,
optimization problems, analytical illustrations, and numerical results
are provided to explain the main design tradeoffs and network-level
insights. We further discuss generalized physical realizations and
representative deployment modes, including standalone, conventional
base-station-integrated, and distributed deployments, together with
their main implementation considerations. Finally, we identify future
research directions involving mobility management, programmable and
virtualized radio access, environment knowledge acquisition,
AI-native control, low-altitude networks, space--air--ground
integration, and integrated sensing and communication. Overall, this
tutorial presents radiation locations as configurable network
resources and provides a framework for understanding how pinching
antennas may reshape future wireless network design and operation.
\end{abstract}


\section{Introduction}

\subsection{Evolution of Wireless Network Architectures}

The evolution of mobile communications from the second generation (2G) to
the fifth generation (5G), and now toward sixth-generation (6G) networks,
has been accompanied not only by advances in air-interface technologies,
but also by continuous changes in wireless network architecture
\cite{you2021towards,wang2023road,tataria20216g}. Each generation has introduced new
requirements in terms of coverage, capacity, latency, reliability,
connectivity density, and service diversity. In response, wireless networks
have progressively evolved from conventional macro-cellular deployments
toward more coordinated, dense, distributed, and user-centric forms of
radio access \cite{3gpp_tr36819,checko2014cloud}. A common motivation behind
this evolution has been to reduce the effective communication distance,
improve spatial reuse and service uniformity, and make network resources
more adaptive to heterogeneous user distributions and traffic demands.

Early cellular systems, including 2G and the initial stages of third
generation (3G), were largely built around a conventional macro-cellular
architecture \cite{rappaport2002wireless}. A geographical region was
partitioned into cells, each served by a fixed base station (BS) responsible
for radio access within its coverage area. This architecture provided an
effective means of achieving wide-area coverage, frequency reuse, and
manageable network planning, and the cell consequently became the basic
unit of wireless network organization. However, as mobile communication
evolved from voice-dominated services toward broadband data, the
limitations of this architecture became increasingly apparent. Users near
cell boundaries suffered from large propagation losses and inter-cell
interference, while spatially nonuniform traffic could lead to substantial
load imbalance among neighboring cells.

The evolution toward 4G and Long-Term Evolution (LTE)-Advanced therefore
brought increasing attention to cooperation among cells and transmission
points. Coordinated multipoint (CoMP) transmission and reception,
coordinated scheduling, and inter-cell interference coordination were
developed to improve cell-edge performance by allowing multiple
geographically separated transmission points to coordinate their operation
\cite{gesbert2010multi,lee2012coordinated}. Such techniques represented an
important departure from purely independent-cell operation by enabling
neighboring cells to coordinate their transmission and reception strategies
rather than operating as isolated coverage units. Nevertheless, such
cooperation was still performed among transmission points whose physical
locations were predetermined by deployment, while the corresponding
performance gains required additional channel information exchange,
synchronization, and backhaul support.

In parallel, network densification emerged as another major architectural
direction. Small cells, pico cells, femtocells, and heterogeneous networks
introduced low-power access nodes within or alongside conventional
macro-cellular coverage, thereby shortening the distance between
transmitters and users and increasing spatial reuse
\cite{andrews2012femtocells,andrews2013seven}. This development became
particularly important toward 4G and 5G, where rapidly increasing
mobile-data demand motivated much denser spatial reuse of spectrum. Indeed,
network densification was widely recognized as one of the major approaches
for increasing cellular capacity toward 5G \cite{andrews2014what}. At the
same time, dense and heterogeneous deployments introduced new challenges
in interference management, user association, mobility, load balancing,
backhaul provisioning, and network planning \cite{andrews2014overview}.
Thus, bringing more fixed access points closer to users improved spatial
service capability, but also substantially increased the complexity of
network deployment and operation.

Another important architectural evolution has been the decoupling of
distributed radio access from centralized or cloud-based processing.
Cloud radio access networks (C-RANs), for example, centralize baseband
processing in shared processing pools while retaining geographically
distributed radio units \cite{china_mobile_cran,checko2014cloud}. Fog radio
access networks (F-RANs) further introduce processing and storage
capabilities closer to the network edge, providing additional flexibility
in balancing centralized coordination and local processing
\cite{peng2014heterogeneous,peng2016fog}. The 5G new-radio (NR)
architecture further supports flexible functional separation of radio
access network components, including centralized and distributed units,
providing greater flexibility in how radio access functions are deployed
and coordinated \cite{3gpp_ts38300}. These developments reflect a broader
transition in which signal processing and network control become
increasingly centralized, distributed, or virtualized according to service
requirements, while radio transmission remains geographically distributed.

More recently, distributed and cell-free architectures have further
challenged the conventional notion of fixed cells. In cell-free massive
multiple-input multiple-output (MIMO), a large number of geographically
distributed access points jointly serve users over the same time-frequency
resources, with the objective of providing more uniform service and
reducing the distinction between cell-center and cell-edge users
\cite{xu2025distributed,ngo2017cell,bjornson2020scalable}. Rather than associating each user
with a single predefined cell, distributed access points can cooperate
according to user-specific channel conditions, leading naturally toward
user-centric network operation. The seminal cell-free massive MIMO
framework demonstrated substantial gains in uniformly achievable user
rates compared with conventional small-cell operation, illustrating the
value of combining distributed service points with coordinated processing
\cite{ngo2017cell}.

Accordingly, the evolution from macro-cellular systems to coordinated
multi-cell networks, heterogeneous and dense deployments, C-RAN/F-RAN,
and cell-free architectures reflects a progressive transition from rigid
cell-centric operation toward more flexible, distributed, and user-centric
network organization. Wireless networks have evolved from serving users
through geographically defined cells to coordinating multiple cells,
deploying increasingly dense service points, centralizing or virtualizing
processing, and eventually allowing distributed access points to jointly
provide user-centric service. This trend is consistent with the broader
evolution toward 5G and beyond, where network flexibility, densification,
distributed access, and adaptive resource management have become
important design themes \cite{dahlman20205g}.

Looking toward 6G, this architectural evolution is expected to continue.
Current 6G visions emphasize increasingly intelligent, distributed, and
heterogeneous networks that integrate communication with sensing and
computing while supporting diverse applications and dynamic environments
\cite{saad2020vision,zhang20196g}. Emerging discussions on future RAN
architectures further highlight the importance of flexibility,
programmability, and functional adaptation in future wireless networks
\cite{dang2020should,jiang2021road}.

\subsection{The Remaining Spatial Limitation of Existing Wireless Architectures}

Despite the substantial architectural advances discussed above, existing
wireless networks still share a fundamental limitation. While network
operation has become increasingly flexible through coordination, resource
allocation, and processing, the physical locations where electromagnetic
waves are radiated into the propagation environment remain largely fixed
after deployment. This limitation persists across different architectural
paradigms. Macro BSs, small cells, remote radio heads, and distributed
access points may differ significantly in their density, coordination
capability, and processing architecture, but their radiation sites are
generally determined during network deployment. Consequently, existing
networks mainly adapt wireless service through user association,
scheduling, power allocation, beamforming, processing coordination, and
cooperation among predefined transmission points.

As a result, current architectures can optimize communication over a given
spatial radio infrastructure, but have limited capability to control where
wireless service is provided. User distribution, traffic demand, and
propagation conditions such as blockage and visibility are therefore
mainly accommodated through transmission and network optimization rather
than treated as controllable design factors. As future networks become
increasingly adaptive and environment-dependent, such fixed radiation
infrastructure may become an important constraint. This motivates the
exploration of new wireless architectures that provide additional
flexibility in determining where wireless signals are radiated and where
wireless service is provided.

This naturally leads to a fundamental question:

\begin{tcolorbox}[colback=green!9!white,colframe=black!35!white,
boxrule=0.5pt,arc=1mm,left=1mm,right=1mm,top=1mm,bottom=1mm]
\emph{Can future wireless architectures reconfigure not only transmission
parameters and network operations, but also the physical locations where
wireless signals are radiated after deployment?}
\end{tcolorbox}

\subsection{Pinching Antennas as a Promising Answer}

Pinching antennas, originally proposed by NTT DOCOMO, introduce a new
paradigm for wireless transmission by enabling electromagnetic waves to be
radiated at configurable locations along a guiding structure
\cite{suzuki2022pinching}. The key idea is to convey radio-frequency (RF)
signals through a guided medium, such as a dielectric waveguide, and
generate radiation at desired locations along the waveguide. By
configuring these radiation points after deployment, pinching antennas
enable radiation-location reconfiguration over potentially large spatial
scales, which is fundamentally different from conventional wireless
infrastructures with fixed radiation sites \cite{ding2024flexible}. This
capability makes the physical locations where wireless signals enter the
propagation environment a controllable network resource.

This capability also distinguishes pinching antennas from existing
flexible-antenna technologies. Reconfigurable intelligent surfaces (RISs)
and intelligent reflecting surfaces (IRSs) mainly modify the propagation
environment by adjusting their electromagnetic response, such as the
element-wise reflection phases \cite{tang2020wireless,wu2019intelligent}.
Fluid antennas and movable antennas enable antenna-position
reconfiguration, but typically within a compact physical region or a
limited movement range, often on the order of several wavelengths
\cite{wong2020fluid,zhu2023modeling,shao20246d}. In contrast, pinching
antennas exploit guided signal propagation along extended structures to
enable radiation-location reconfiguration over a much larger spatial
domain. They therefore introduce a distinct form of spatial flexibility
that extends beyond local antenna reconfiguration and conventional
transmission optimization over fixed infrastructure.

\begin{figure*}[!t]
\centering
\includegraphics[width=0.98\linewidth]{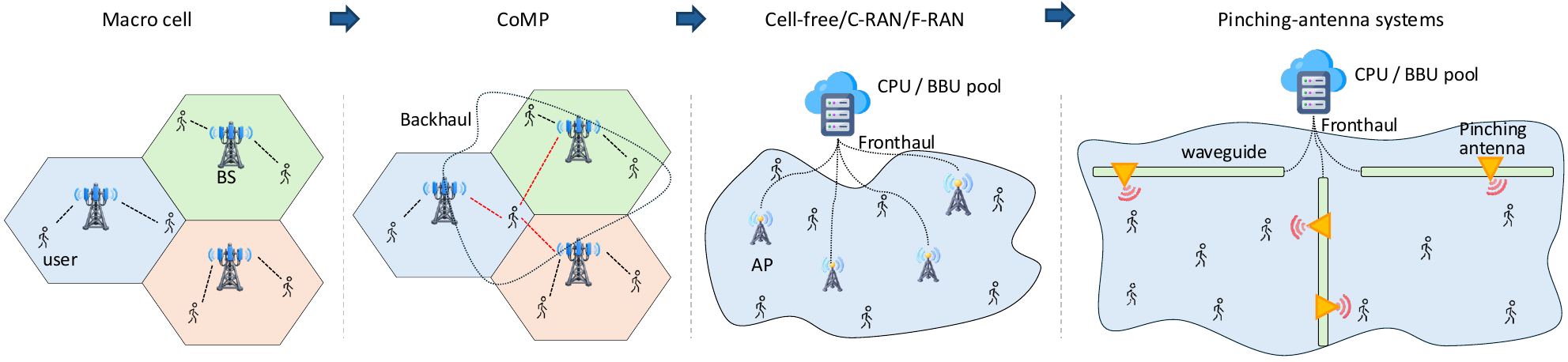}
\captionsetup{justification=justified, singlelinecheck=false, font=small}
\caption{Evolution of representative wireless network architectures from macro-cellular networks to pinching-antenna systems.}
\label{fig: fixed to pa}
\end{figure*}

\begin{table*}[!t]
\centering
\caption{Comparison of representative wireless network architectures.}
\renewcommand{\arraystretch}{1.5}
\setlength{\tabcolsep}{2pt}

\begin{tabularx}
{\linewidth}
{
p{2.6cm}
p{3.2cm}
p{3.4cm}
p{4.2cm}
X
}
\hline \hline

{\centering\textbf{Aspect}}
& \textbf{Macro cell}
& \textbf{CoMP}
& \textbf{Cell-free / C-RAN / F-RAN}
& \textbf{Pinching-antenna systems} \\
\hline

\multirow{1.6}{*}{{\cellcolor{gray!12}\textbf{Network organization}}}
& \multirow{1.6}{*}{{\cellcolor{gray!12}Cell-centric}}
& \multirow{1.6}{*}{{\cellcolor{gray!12}Coordinated multi-cell}}
& \multirow{1.6}{*}{{\cellcolor{gray!12}Distributed / user-centric}}
& {\cellcolor{gray!12}Distributed with reconfigurable radiation locations} \\

{\textbf{Radiation location}}
& Fixed BS locations
& Fixed BS locations
& Fixed AP / RRH locations
& Reconfigurable along waveguides \\

\multirow{1.6}{*}{{\cellcolor{gray!12}\textbf{Coordination}}}
& \multirow{1.6}{*}{{\cellcolor{gray!12}Mainly within each cell}}
& \multirow{1.6}{*}{{\cellcolor{gray!12}Among neighboring BSs}}
& \multirow{1.6}{*}{{\cellcolor{gray!12}Among distributed APs / RRHs}}
& {\cellcolor{gray!12}Joint transmission and radiation-location control} \\

{\textbf{Cell boundary / service region}}
& \multirow{1.6}{*}{Fixed}
& \multirow{1.6}{*}{Largely fixed}
& \multirow{1.6}{*}{Weakened or removed}
& \multirow{1.6}{*}{Reconfigurable after deployment} \\

\hline \hline
\end{tabularx}

\label{tab:architecture_comparison}
\end{table*}

From a network architecture perspective, making radiation locations
controllable over large spatial scales introduces a new degree of freedom
that can enable several new network functions. First, by changing the
locations from which signals are radiated, pinching antennas can modify
desired and interfering links, creating new opportunities for user
separation, multiple access, and interference management. Second,
radiation-location control can change the effective serving relationships
between transmission points and users, providing additional flexibility
for cell association, multi-cell coordination, and traffic offloading.
Third, distributed waveguides with reconfigurable radiation points allow
wireless coverage to be adapted according to user distribution, traffic
demand, and environmental conditions, enabling more flexible region-level
service design. Beyond these examples, the separation between processing
functions and configurable radiation locations can also provide additional
flexibility for distributed network architectures.

\newcolumntype{C}[1]{%
    >{\centering\arraybackslash}m{#1}%
}

\begin{table*}[!t]
    \centering
    \caption{Comparison of this paper with representative existing
    works on pinching-antenna systems.}
    \label{tab:comparison_existing_pa_works}

    \scriptsize
    \setlength{\tabcolsep}{2pt}
    \renewcommand{\arraystretch}{1.85}

    \begin{tabular}{|
        C{0.075\textwidth}|
        C{0.075\textwidth}|
        C{0.105\textwidth}|
        C{0.125\textwidth}|
        C{0.060\textwidth}|
        C{0.07\textwidth}|
        C{0.08\textwidth}|
        C{0.095\textwidth}|
        C{0.105\textwidth}|
        C{0.125\textwidth}|}
        \hline

        \multirow{2}{*}{Reference}
        &
        \multirow{2}{*}{Type}
        &
        \multirow{2}{*}{%
            \makecell{Principles and\\channel modeling}}
        &
        \multirow{2}{*}{%
            \makecell{Physical realizations\\and hardware\\architectures}}
        &
        \multicolumn{3}{c|}{Higher-layer design}
        &
        \multicolumn{2}{c|}{Physical-layer design}
        &
        \multirow{2}{*}{%
            \makecell{Network realizations\\and deployment\\modes}}
        \\
        \cline{5-9}

        &
        &
        &
        &
        EDMA
        &
        \makecell{Multi-cell\\design}
        &
        \makecell{Region-level\\design}
        &
        \makecell{Communication\\design}
        &
        \makecell{Multifunctional\\applications}
        &
        \\
        \hline

        \cite{ding2024flexible}
        & Technical
        & \fullcircle
        & \partialcircle
        & \emptycircle
        & \emptycircle
        & \emptycircle
        & \fullcircle
        & \emptycircle
        & \emptycircle
        \\
        \hline

        \cite{yang2025pinching}
        & Magazine
        & \fullcircle
        & \partialcircle
        & \emptycircle
        & \emptycircle
        & \emptycircle
        & \fullcircle
        & \fullcircle
        & \emptycircle
        \\
        \hline

        \cite{liu2026architecture}
        & Magazine
        & \fullcircle
        & \partialcircle
        & \emptycircle
        & \emptycircle
        & \emptycircle
        & \fullcircle
        & \fullcircle
        & \emptycircle
        \\
        \hline

        \cite{lu2026survey}
        & Survey
        & \fullcircle
        & \fullcircle
        & \emptycircle
        & \emptycircle
        & \emptycircle
        & \fullcircle
        & \fullcircle
        & \partialcircle
        \\
        \hline

        \cite{liu2026survey}
        & Survey
        & \fullcircle
        & \fullcircle
        & \emptycircle
        & \emptycircle
        & \emptycircle
        & \fullcircle
        & \fullcircle
        & \emptycircle
        \\
        \hline

        \cite{illi2026survey}
        & Survey
        & \fullcircle
        & \fullcircle
        & \emptycircle
        & \emptycircle
        & \emptycircle
        & \fullcircle
        & \fullcircle
        & \partialcircle
        \\
        \hline

        \cite{xu2026generalized}
        & Tutorial
        & \fullcircle
        & \fullcircle
        & \emptycircle
        & \emptycircle
        & \emptycircle
        & \fullcircle
        & \fullcircle
        & \partialcircle
        \\
        \hline

        \cite{liu2026tutorial}
        & Tutorial
        & \fullcircle
        & \fullcircle
        & \emptycircle
        & \emptycircle
        & \emptycircle
        & \fullcircle
        & \fullcircle
        & \emptycircle
        \\
        \hline

        This paper
        & Tutorial
        & \fullcircle
        & \fullcircle
        & \fullcircle
        & \fullcircle
        & \fullcircle
        & \partialcircle
        & \partialcircle
        & \fullcircle
        \\
        \hline
    \end{tabular}

    \vspace{1.5mm}

    \parbox{\textwidth}{%
        \scriptsize
        \textrm {Note:}
        ``\fullcircle'', ``\partialcircle'', and ``\emptycircle'' indicate
        a key focus, partial discussion, and no discussion,
        respectively. EDMA denotes environment-division multiple
        access.
    }
\end{table*}

\begin{figure*}[ht]
\centering
\includegraphics[height=0.72\textwidth]{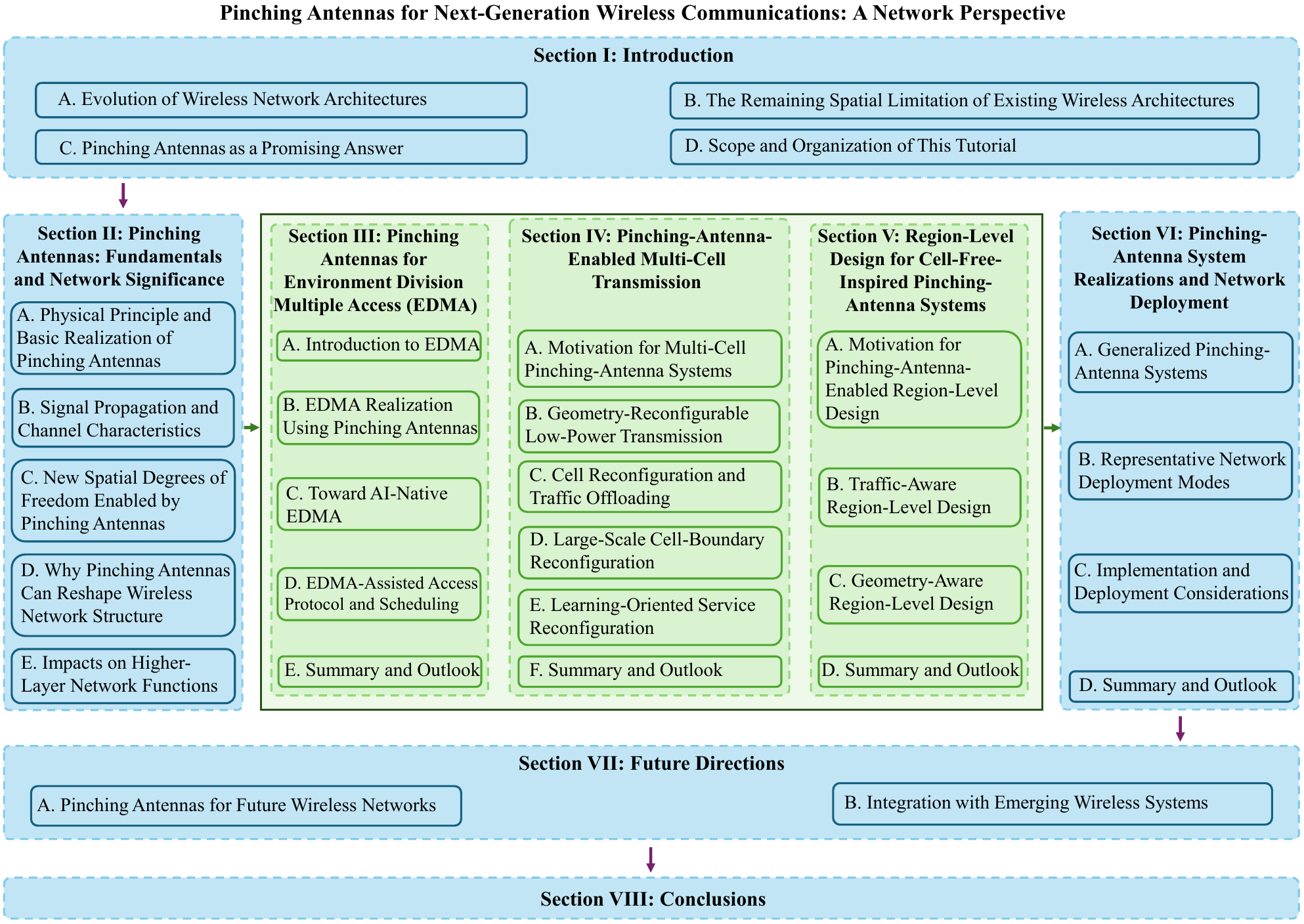}
\captionsetup{justification=justified, singlelinecheck=false, font=small}
\caption{The condensed structure of this tutorial.}
\label{fig: outline}
\end{figure*}

In summary, pinching antennas represent more than a new antenna
technology. By transforming radiation locations from fixed deployment
parameters into controllable network resources, they extend wireless
network design from optimizing transmission and coordination over fixed
radio sites to also controlling where wireless service is provided.
Fig. \ref{fig: fixed to pa} illustrates this evolution from conventional cell-centric
networks toward pinching-antenna systems with reconfigurable radiation
locations, while Table \ref{tab:architecture_comparison} further summarizes their main differences in
network organization, radiation locations, coordination, and service
regions. Building on this new degree of freedom, the remainder of this paper
investigates how radiation-location reconfigurability can support new
network functions.

\subsection{Scope and Organization of This Tutorial}

Several magazine articles, surveys, and tutorials on pinching-antenna
systems have recently appeared in the literature. These works have
mainly discussed operating principles, channel models, physical
realizations, hardware architectures, physical-layer transmission
design, and multifunctional applications, while also providing
broader reviews of optimization and intelligent design methods
\cite{yang2025pinching,liu2026architecture,lu2026survey,
liu2026survey,illi2026survey,xu2026generalized,liu2026tutorial}.
Complementing the existing literature, this paper focuses on the
implications of pinching antennas for wireless network design and
operation. In particular, we treat configurable radiation locations
as network resources and investigate how they can reshape
interference, serving relationships, region-level service provision,
and network deployment. A comparison of the scope of this paper with
representative existing works is provided in
Table~\ref{tab:comparison_existing_pa_works}.

The main contributions of this paper are summarized as follows:
\begin{itemize}
    \item We introduce the architecture and channel characteristics of
    pinching-antenna systems from a network perspective. Based on
    radiation-location reconfigurability, we explain how pinching antennas can modify
    desired and interfering links, change the effective relationships
    between transmission points and users, and adapt wireless service
    across a communication region.

    \item We investigate the roles of pinching antennas in multiple-access and
    multi-cell networks. Environment-division multiple access (EDMA) is
    presented as a representative framework for exploiting differences
    between desired and cross-links, while multi-cell operation is
    discussed in the contexts of cell association, traffic offloading,
    inter-cell interference management, and cooperative transmission.
    Representative analytical examples, optimization problems, and
    numerical results are provided to illustrate the main design
    tradeoffs.

    \item We extend pinching-antenna design from instantaneous users to entire
    communication regions under a cell-free-inspired multi-waveguide
    architecture. Traffic-aware and geometry-aware designs are
    introduced to demonstrate how long-term spatial demand and
    environmental blockage information can guide pinching-antenna configuration for
    improving network-average performance and region-wide coverage.

    \item We discuss representative physical realizations and network
    deployment modes, including standalone, conventional
    BS-integrated, and distributed deployments. We further identify
    future directions concerning mobility management, programmable
    radio access, environment knowledge acquisition, AI-native
    operation, and the integration of pinching antennas with selected emerging
    wireless technologies.
\end{itemize}

The organization of this paper is illustrated in
Fig.~\ref{fig: outline}. The remainder of the paper is organized as
follows. Sec.~\ref{sec: PA fundamental} introduces the
pinching-antenna system architecture and channel models and explains
the network-level significance of radiation-location
reconfigurability. Sec.~\ref{sec: edma} discusses how pinching
antennas can support EDMA. Sec.~\ref{sec: multicell} examines the
roles of pinching antennas in multi-cell networks. Sec.~\ref{sec: region level}
presents region-level design under a cell-free-inspired architecture,
using traffic-aware and geometry-aware designs as representative
examples. Sec.~\ref{sec: realization and deployment} introduces
generalized pinching-antenna realizations and representative network deployment
modes and discusses the associated implementation considerations.
Sec.~\ref{sec: future direction} outlines future research directions for
pinching-antenna-enabled wireless networks and their integration with emerging
wireless technologies. Finally, Sec.~\ref{sec: conclusion} concludes
the paper.
Table \ref{tab:acronyms} summarizes the main acronyms used throughout the paper.

\begin{table*}[!t]
\centering
\caption{List of acronyms used in this paper.}
\label{tab:acronyms}
\setlength{\tabcolsep}{5pt}
\renewcommand{\arraystretch}{1.25}
\footnotesize
\begin{tabular}{
    |>{\centering\arraybackslash}m{0.075\textwidth}
    |>{\raggedright\arraybackslash}m{0.37\textwidth}
    |>{\centering\arraybackslash}m{0.075\textwidth}
    |>{\raggedright\arraybackslash}m{0.37\textwidth}|
}
\hline
\textbf{Acronym} & \textbf{Definition}
& \textbf{Acronym} & \textbf{Definition} \\
\hline
2G    & Second generation
& 3G    & Third generation \\
\hline
4G    & Fourth generation
& 5G    & Fifth generation \\
\hline
6G    & Sixth generation
& AI    & Artificial intelligence \\
\hline
AP    & Access point
& APU   & Antenna processing unit \\
\hline
BBU   & Baseband unit
& BS    & Base station \\
\hline
CCDF  & Complementary cumulative distribution function
& CDMA  & Code-division multiple access \\
\hline
CoMP  & Coordinated multipoint
& CPU   & Central processing unit \\
\hline
C-RAN & Cloud radio access network
& EDMA  & Environment-division multiple access \\
\hline
FDMA  & Frequency-division multiple access
& FL    & Federated learning \\
\hline
F-RAN & Fog radio access network
& IID   & Independent and identically distributed \\
\hline
IRS   & Intelligent reflecting surface
& ISAC  & Integrated sensing and communication \\
\hline
LCX   & Leaky coaxial cable
& LEO   & Low Earth orbit \\
\hline
LoS   & Line-of-sight
& LTE   & Long-Term Evolution \\
\hline
MAC   & Medium access control
& MIMO  & Multiple-input multiple-output \\
\hline
MRT   & Maximum-ratio transmission
& NLoS  & Non-line-of-sight \\
\hline
NOMA  & Non-orthogonal multiple access
& NR    & New radio \\
\hline
PDCP  & Packet data convergence protocol
& PEMNet & Perception-embedding-map-enabled mobile network \\
\hline
PHY   & Physical layer
& QoS   & Quality of service \\
\hline
RAN   & Radio access network
& RF    & Radio frequency \\
\hline
RIS   & Reconfigurable intelligent surface
& RLC   & Radio link control \\
\hline
RRC   & Radio resource control
& RRH   & Remote radio head \\
\hline
RS    & Radio stripe
& SCA   & Successive convex approximation \\
\hline
SDAP  & Service data adaptation protocol
& SDMA  & Space-division multiple access \\
\hline
SIC   & Successive interference cancellation
& SINR  & Signal-to-interference-plus-noise ratio \\
\hline
SNR   & Signal-to-noise ratio
& TDMA  & Time-division multiple access \\
\hline
UAV   & Unmanned aerial vehicle
& URLLC & Ultra-reliable low-latency communications \\
\hline
\end{tabular}
\end{table*}

\textbf{Notations:} Throughout this paper, scalars are denoted by
italic letters, while lowercase and uppercase bold letters, such as
$\boldsymbol{x}$ and $\boldsymbol{X}$, denote column vectors and
matrices, respectively. The sets of real and complex
$M\times N$ matrices are denoted by $\mathbb{R}^{M\times N}$ and
$\mathbb{C}^{M\times N}$, respectively. The superscripts
$(\cdot)^{\mathsf T}$ and $(\cdot)^{*}$ denote the transpose and
complex conjugate, respectively, and $j=\sqrt{-1}$ is the imaginary
unit. The notation $\lVert\boldsymbol{x}\rVert$ denotes the Euclidean
norm of $\boldsymbol{x}$, while $|x|$ denotes the absolute value of a
real scalar or the modulus of a complex scalar. For a finite set
$\mathcal{S}$, $|\mathcal{S}|$ denotes its cardinality.
The operators $\mathbb{E}[\cdot]$ and $\Pr\{\cdot\}$ denote
expectation and probability, respectively; a subscript on
$\mathbb{E}$ specifies the random variables over which the expectation
is taken. The notation
$\mathcal{N}(\boldsymbol{x};\boldsymbol{\mu},\boldsymbol{\Sigma})$
denotes the probability density of a Gaussian random vector with mean
$\boldsymbol{\mu}$ and covariance matrix $\boldsymbol{\Sigma}$,
whereas $x\sim\mathcal{CN}(\mu,\sigma^2)$ denotes a circularly
symmetric complex Gaussian random variable with mean $\mu$ and
variance $\sigma^2$. Finally, $\mathbbm{1}\{\mathcal{A}\}$ denotes the
indicator of event $\mathcal{A}$, and
$\operatorname{diag}(\cdot)$ constructs a diagonal matrix from its
arguments.

\section{Pinching Antennas: Fundamentals and Network Significance} \label{sec: PA fundamental}
This section introduces the fundamentals of pinching-antenna systems and
establishes their significance from a network perspective. We first
describe the physical realization and channel characteristics of
pinching antennas, and then explain the new spatial degree of freedom
enabled by radiation-location reconfigurability. Based on these
fundamentals, we further discuss why this capability can reshape wireless
network architectures and affect higher-layer network functions.

\subsection{Physical Principle and Basic Realization of Pinching Antennas}

\begin{figure}[!t]
    \centering
    \includegraphics[width=0.92\linewidth]{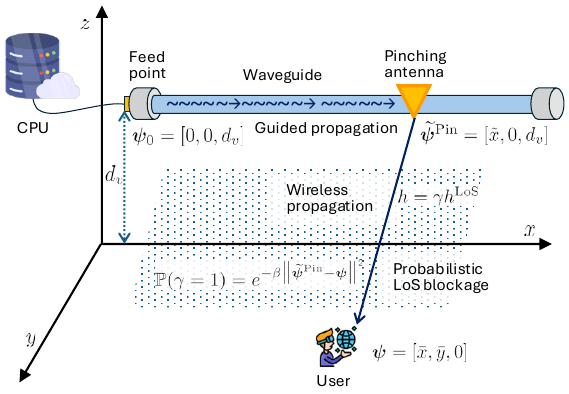}
    \captionsetup{justification=justified, singlelinecheck=false, font=small}
    \caption{Illustration of the physical realization and channel characteristics of a pinching-antenna system, including guided propagation along the waveguide, wireless propagation from the pinching antenna to the user, and distance-dependent probabilistic LoS blockage.}
\label{fig: pa model}
\end{figure}

A canonical pinching-antenna system consists of a CPU for baseband signal
processing, a dielectric waveguide for RF signal delivery, and one or more
pinching antennas for forming configurable radiation points, as illustrated
in Fig.~\ref{fig: pa model}. Taking downlink transmission as an example,
the processed signal is converted to RF and fed into the dielectric
waveguide, where it propagates as a guided electromagnetic wave. A
pinching antenna placed at a selected position couples the guided
signal into free space, thereby forming an effective radiation point for
wireless transmission \cite{suzuki2022pinching,ding2024flexible}. The
transmitted signal therefore undergoes two consecutive propagation stages,
namely guided propagation along the waveguide and wireless propagation
from the pinching antenna to the user.

An important feature of this realization is the separation between signal
delivery and wireless radiation. The waveguide provides an extended
physical path for transporting the RF signal, whereas the pinching antenna
determines where the signal is radiated into the surrounding environment.
By configuring the pinching antenna at different positions along the
waveguide, the radiation location can be changed without relocating the
RF source or the underlying waveguide infrastructure. Multiple pinching
antennas can also be configured on the same waveguide when multiple
radiation points are required. Consequently, the same deployed waveguide
can support different radiation configurations according to communication
requirements.

This guided-then-radiated mechanism is fundamentally different from
conventional fixed-position antenna systems, where signal feeding and
wireless radiation are tied to predetermined antenna locations. For
pinching antennas, the location at which the signal enters the wireless
propagation environment becomes configurable after deployment. This
physical property gives rise to distinctive location-dependent channel
characteristics, which are discussed next.

\subsection{Signal Propagation and Channel Characteristics} \label{subsec: channel model}

Building on the guided-then-radiated transmission mechanism described
above, we next examine the basic channel characteristics of
pinching-antenna systems. For clarity, we consider a single-user,
single-waveguide setting, as illustrated in
Fig.~\ref{fig: pa model}. The user is located at
$\boldsymbol{\psi}=[\bar{x},\bar{y},0]$, while the waveguide is deployed
parallel to the $x$-axis at height $d_v$, with its feed point located at
$\boldsymbol{\psi}_0=[0,0,d_v]$. The location of the pinching antenna is
denoted by
$\widetilde{\boldsymbol{\psi}}^{\rm Pin}=[\tilde{x},0,d_v]$, where
$\tilde{x}$ determines the radiation location along the waveguide.

We first consider the basic case where a line-of-sight (LoS) link is established
between the pinching antenna and the user. As illustrated in Fig. \ref{fig: pa model}, the signal experiences guided
propagation from the feed point to the pinching antenna, followed by
free-space propagation from the pinching antenna to the user. Accordingly,
the equivalent LoS channel between the feed point and the user can be
expressed as 
\begin{align}
h^{\rm LoS}
=
\frac{\eta^{\frac{1}{2}} \exp\left(
-j\left[
\frac{2\pi}{\lambda}
\|\boldsymbol{\psi}-\widetilde{\boldsymbol{\psi}}^{\rm Pin}\|
+
\frac{2\pi}{\lambda_g}
\|\boldsymbol{\psi}_0-\widetilde{\boldsymbol{\psi}}^{\rm Pin}\|
\right]
\right)}
{\|\boldsymbol{\psi}-\widetilde{\boldsymbol{\psi}}^{\rm Pin}\|}
,
\label{eq:pa_los_channel}
\end{align}
where $\eta=\frac{\lambda^2}{16\pi^2}$ denotes the reference free-space
channel gain, $\lambda$ is the free-space wavelength, and
$\lambda_g=\frac{\lambda}{n_{\rm eff}}$ denotes the guided wavelength,
where $n_{\rm eff}$ is the effective refractive index of the dielectric
waveguide \cite{ding2024flexible}.

The channel model in \eqref{eq:pa_los_channel} exhibits two distinctive
characteristics. First, the channel amplitude depends on the free-space
distance between the configurable radiation location and the user.
Therefore, changing the location of a pinching antenna along the waveguide can
substantially change the propagation distance and the corresponding
large-scale channel gain. Second, the overall channel phase contains
two components, one introduced by free-space propagation and the other
by guided propagation along the waveguide. Since both propagation
distances depend on $\tilde{x}$, reconfiguring the radiation location
also changes the channel phase. These characteristics distinguish
pinching-antenna channels from conventional fixed-site channels, where
the radiation location is predetermined by infrastructure deployment.

In practical propagation environments, it may contain obstacles that make LoS availability
location dependent. To capture this effect in a simple form, let
$\gamma\in\{0,1\}$ denote the LoS indicator, where $\gamma=1$
indicates that an LoS path is available. Following a distance-dependent
probabilistic LoS model \cite{bai2014analysis}, the LoS probability can be expressed as
\begin{align}
\Pr(\gamma=1)
=
\exp\left(-\beta\|\widetilde{\boldsymbol{\psi}}^{\rm Pin}-\boldsymbol{\psi}\|^2
\right),
\label{eq:pa_los_probability}
\end{align}
where $\beta>0$ characterizes the blockage density of the propagation
environment. A larger $\beta$ corresponds to a lower probability of
maintaining an LoS link for a given propagation distance \cite{ding2025blockage,xu2025pinching-los,li2026power,xu2026_losnlos}. The resulting
channel can then be represented as
\begin{align}
h=\gamma h^{\rm LoS}.
\label{eq:pa_prob_los_channel}
\end{align}

The probabilistic LoS model reveals another important effect of
radiation-location reconfigurability. In addition to changing the
free-space propagation distance and channel phase, configuring the
pinching-antenna location can also change the likelihood of establishing
an LoS connection. For example, the average channel power under
\eqref{eq:pa_los_probability} is given by
\begin{align}
\mathbb{E}_{\gamma}\left[|h|^2\right]
=
\frac{\eta\exp\left(-\beta\|\widetilde{\boldsymbol{\psi}}^{\rm Pin}-\boldsymbol{\psi}\|^2
\right)}
{\|\widetilde{\boldsymbol{\psi}}^{\rm Pin}-\boldsymbol{\psi}
\|^2}.
\label{eq:pa_average_channel_gain}
\end{align}
Hence, the radiation location jointly determines the free-space path
loss and the LoS availability. A suitable radiation location can not
only shorten the wireless propagation distance, but also increase the
probability of maintaining a favorable LoS connection in blockage-prone
environments.

Overall, the channel characteristics of pinching antennas are closely
coupled with their configurable radiation locations. Reconfiguring the
pinching antenna changes the free-space propagation distance, the
propagation phase, and, under probabilistic blockage, the LoS
availability. These location-dependent channel characteristics provide
the physical basis for the new spatial degree of freedom discussed in
the following subsection.

\begin{figure*}[!t]
    \centering
    \includegraphics[width=0.98\linewidth]{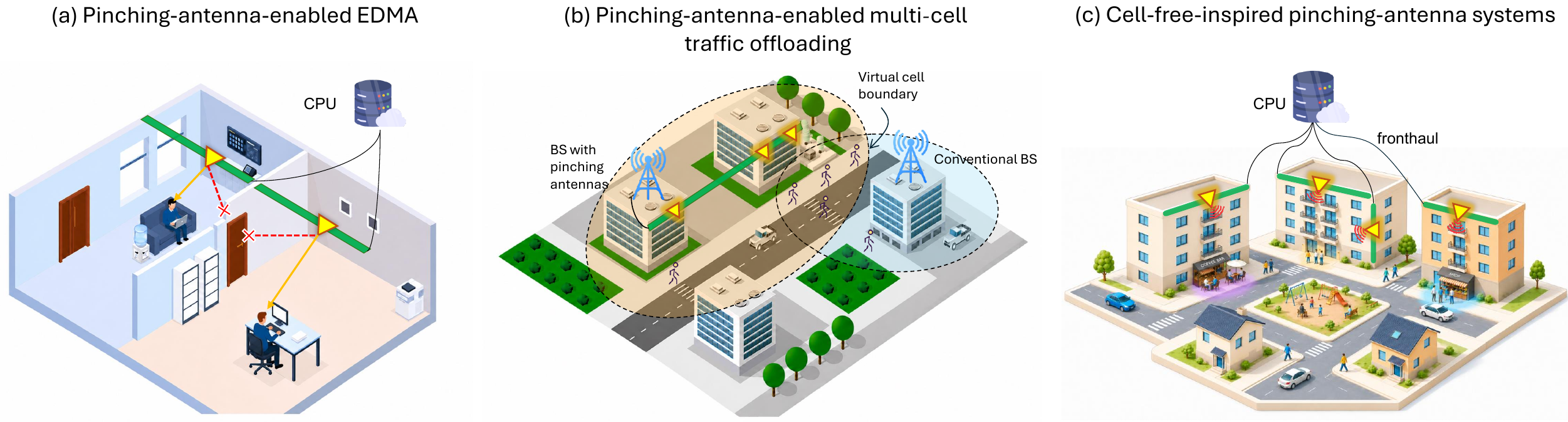}
    \captionsetup{justification=justified, singlelinecheck=false, font=small}
    \caption{Illustration of how pinching antennas can reshape wireless
    network structure. (a) Pinching-antenna-enabled EDMA reconfigures desired and
    interfering links by adjusting radiation locations. (b) Pinching-antenna-enabled
    multi-cell traffic offloading changes serving relationships and creates
    adaptive virtual cell boundaries. (c) A cell-free-inspired pinching-antenna
    architecture connects distributed waveguides to a common processing
    unit, enabling coordinated region-wide service without rigid cell
    boundaries.}
\label{fig: pa three cases}
\end{figure*}

\subsection{New Spatial Degrees of Freedom Enabled by Pinching Antennas} \label{subsec: spatial dof}

In conventional wireless infrastructures, radiation locations are
generally fixed during deployment. System adaptation is therefore mainly
achieved by adjusting transmission and resource-allocation variables,
such as transmit power, beamforming, user scheduling, and spectrum
allocation, over a predetermined radio geometry. Pinching antennas
introduce an additional spatial degree of freedom by allowing the
radiation location to be reconfigured after deployment. For the
single-waveguide example considered above, this degree of freedom is
represented by the controllable location
$\widetilde{\boldsymbol{\psi}}^{\rm Pin}$ along the waveguide.
Changing this location directly modifies the geometry of the wireless
link and, consequently, its propagation characteristics. Wireless
performance can therefore be adapted not only through signal processing
and resource allocation, but also by controlling where the signal is
radiated into the wireless environment.

Another important feature is its spatial scale. Unlike conventional
flexible-antenna techniques, where antenna locations are typically
adjusted within a relatively compact region, a pinching antenna can in
principle be configured over an extended waveguide whose length is
determined by the deployment environment. The resulting reconfiguration
range can therefore be much larger than the wavelength-scale range
commonly considered in conventional antenna-location optimization. When
multiple waveguides or multiple pinching antennas are employed, this
capability further extends from controlling a single radiation location
to reconfiguring a distributed set of radiation locations across a
communication area.

As discussed in the next subsection, this additional spatial degree of
freedom has implications beyond individual link optimization and can
reshape interference relationships among users, serving relationships
between transmission points and users, and the spatial distribution of
wireless service over a communication region.

\subsection{Why Pinching Antennas Can Reshape Wireless Network Structure}

We next discuss how radiation-location reconfigurability can reshape
wireless network operation and structure. Its architectural implications
can be understood at three progressively broader levels, namely the
multiuser, multi-cell, and region levels. An illustrative overview of
these three levels is provided in Fig. \ref{fig: pa three cases}.

\subsubsection{Reconfigurable Interference Relationships and
Environment-Division Multiple Access}

At the multiuser level, radiation locations directly affect the desired
and interfering links of different users. In conventional wireless
systems, transmitter locations are fixed, so the desired and interfering
links experienced by different users are largely determined by the
network deployment and propagation environment. Multiple-access and
interference-management schemes therefore mainly separate users through
conventional resources such as time, frequency, code, power, and spatial
processing. 

Pinching antennas provide an additional mechanism for shaping this
interference structure. By configuring radiation locations according to
user locations and environmental conditions, the network can change the
relative strengths and propagation conditions of desired and interfering
links. For example, a radiation location can be selected to provide a
favorable LoS link to an intended user while the corresponding
interfering link experiences less favorable propagation, as shown in Fig. \ref{fig: pa three cases}(a). Radiation-location
reconfigurability can therefore create additional differences among
simultaneously served users that can be exploited for user separation.

This capability provides the basis for environment-division multiple
access (EDMA), which explicitly exploits differences in the propagation
environment for multiple access \cite{ding2026environment}. In this sense, the environment is no
longer only a condition that multiple-access design needs to accommodate.
Together with configurable radiation locations, it can also be exploited
to shape multiuser interference and create new access opportunities.
Section \ref{sec: edma} develops this idea further and discusses the associated
access and scheduling strategies.

\subsubsection{Reconfigurable Serving Relationships and Multi-Cell
Operation}

The same radiation-location flexibility also affects network operation
at a broader multi-cell level. In conventional cellular networks, user
association, inter-cell coordination, and traffic offloading are
performed among BSs whose radiating sites remain fixed. Although the
serving BS of a user can be changed, the physical locations from which
different BSs provide wireless service are largely determined during
deployment. As a result, serving relationships and cell geometries remain
strongly constrained by the underlying infrastructure.

With pinching antennas, a BS can control not only whether and how it
serves a user, but also where along its deployed waveguide the signal is
radiated. The effective service point of a BS can therefore be moved
closer to a target user or toward a cell boundary without relocating the
BS or the waveguide infrastructure. This additional flexibility allows
the serving relationships between neighboring cells and users to be
adapted through radiation-location control, as shown in Fig. \ref{fig: pa three cases}(b).

Such flexibility can be exploited together with conventional cell
association, traffic offloading, and multi-cell coordination. For
example, radiation-location control can modify desired-link distances and
inter-cell interference conditions, creating new opportunities for
energy-efficient multi-cell transmission \cite{ding2026toward}.
By using this flexibility, an overloaded cell may be assisted by a neighboring cell whose
radiation location is configured closer to the offloaded users \cite{ding2026impact}. 
Multi-cell operation can
therefore evolve from adapting transmission over largely fixed cell
geometries to jointly adapting transmission strategies and radiation
locations. These possibilities are discussed in Sec. \ref{sec: multicell}.

\subsubsection{Reconfigurable Service Regions and Region-Level Design}

When multiple waveguides are distributed across a communication region,
the impact of radiation-location reconfigurability extends further from
individual serving relationships to the spatial organization of wireless
service. A cell-free-inspired architecture provides a natural framework
for such systems. Multiple waveguides can be deployed along available
building or infrastructure surfaces, connected to a common processing
unit, and jointly used to serve the communication region without relying
on rigid cell boundaries, as shown in Fig. \ref{fig: pa three cases}(c). 
Such an architecture combines distributed
wireless access with configurable radiation locations along each
waveguide.

This additional spatial flexibility distinguishes it from conventional
cell-free systems with fixed AP locations. In a conventional cell-free
network, the network can determine which APs serve a user and how their
transmissions are coordinated, but the AP locations remain fixed after
deployment. In a cell-free-inspired pinching-antenna system, the network
can further control where effective radiation points are formed along
the distributed waveguides. Distributed cooperation and
radiation-location reconfigurability can therefore be jointly exploited
to shape wireless service across the communication region.

This capability naturally leads to region-level design. Rather than
configuring radiation locations only for a given set of instantaneous
users, the network can optimize their spatial distribution according to
longer-term information about traffic demand and environmental
conditions. For example, radiation locations can be configured to
strengthen service in traffic hotspots, improve coverage balance, or
avoid regions with unfavorable blockage conditions \cite{xu2026environment1,xu2026environment2}. The design objective
thus extends from improving individual links to shaping service
performance over an entire communication region. Sec. \ref{sec: region level} develops
this region-level perspective in detail.

Taken together, these three levels reveal a progressive architectural
impact of radiation-location reconfigurability. At the multiuser level,
it can reshape interference relationships among users. At the multi-cell
level, it can reshape serving relationships between transmission
infrastructures and users. At the region level, it can reshape the
spatial distribution of wireless service. Pinching antennas therefore
extend network adaptation beyond transmission and resource allocation
over fixed radiating sites to include control over where wireless service
is provided.

\begin{figure*}[!t]
    \centering
    \includegraphics[width=0.92\linewidth]{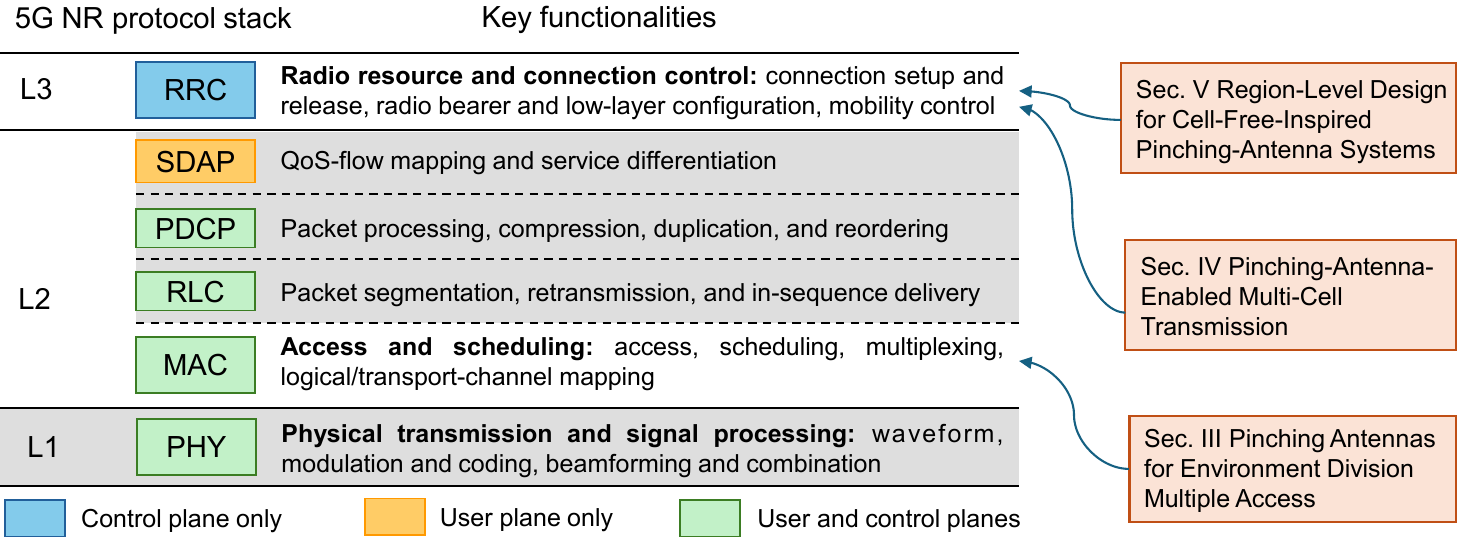}
    \captionsetup{justification=justified, singlelinecheck=false, font=small}
    \caption{Illustration of the 5G NR protocol stack and the network functions considered in this tutorial. The figure provides a functional reference for showing how radiation-location reconfigurability introduced by pinching antennas can affect different levels of wireless network operation. Section \ref{sec: edma} mainly considers MAC-layer multiple-access organization, whereas Sections \ref{sec: multicell}--\ref{sec: region level} investigate progressively broader network functions, including multi-cell coordination, traffic offloading, and region-level service design. PHY is included as the physical foundation enabled by radiation-location reconfigurability.}
\label{fig: protocol sec}
\end{figure*}

\subsection{Impacts on Higher-Layer Network Functions}

These architectural changes also have
important implications for higher-layer network functions. To illustrate
these implications, Fig.~\ref{fig: protocol sec} uses the 5G new radio
(NR) protocol stack as a functional reference \cite{3gpp_ts38300}. At the
physical layer (PHY), radiation-location reconfigurability changes the
wireless channel through the mechanisms discussed in Secs. \ref{subsec: channel model} and \ref{subsec: spatial dof},
providing the physical foundation for higher-layer operation. More
importantly, the resulting changes in channel conditions, interference
relationships, and service availability can affect the information and
conditions on which higher-layer network decisions are made.

The most direct higher-layer impact appears at the medium access control
(MAC) layer. Conventional MAC functions determine how users access the
network and how radio resources are scheduled among them. With pinching
antennas, radiation-location configuration becomes an additional factor
in these decisions. Different radiation locations can create different
desired and interfering channel conditions for a given group of users.
User grouping, multiple access, scheduling, and pinching-antenna activation can
therefore be jointly designed rather than being optimized over a fixed
radio topology. This provides the higher-layer basis for EDMA and creates
new opportunities for access protocols and queue-aware scheduling, in
which radiation configurations can be exploited together with
conventional radio resources.

Radiation-location reconfigurability can also affect radio resource
control (RRC) and broader radio access network (RAN) control functions
that operate over larger spatial and temporal scales. In conventional
cellular networks, serving relationships are mainly adjusted through user
association, handover, and coordination among fixed transmission points.
With pinching antennas, the network can additionally reconfigure the
radiation locations associated with these serving infrastructures.
Consequently, cell association, traffic offloading, multi-cell
coordination, and mobility management can be jointly considered with
radiation-location control. At a broader scale, multiple distributed
waveguides can be configured according to traffic demand and
environmental conditions, enabling region-level service control.

Radiation-location reconfigurability can also affect other higher-layer
functions through the resulting changes in communication performance.
For example, more reliable and stable wireless links can reduce the need
for radio link control (RLC) retransmissions, support more effective
packet data convergence protocol (PDCP) packet duplication and mobility
handling, and help the service data adaptation protocol (SDAP) satisfy
different quality-of-service (QoS) requirements. These effects can be
particularly relevant to demanding services such as ultra-reliable
low-latency communications (URLLC), whose performance depends on
coordinated operation across multiple protocol layers. Radiation-location
configuration does not typically enter these protocol functions as a
direct control variable. Instead, it affects their operation through the
resulting changes in reliability, latency, service continuity, and other
communication conditions. Accordingly, the higher-layer discussion in
this paper focuses mainly on MAC functions and RRC-oriented network
control, where radiation-location reconfigurability can be more directly
integrated with access, scheduling, association, mobility, coordination,
and region-level service design.

These network functions also operate over different time scales.
Radiation-location reconfiguration associated with user access and
scheduling may need to respond relatively quickly, whereas reconfiguration
for mobility management, traffic adaptation, multi-cell coordination, or
region-level service design can generally follow slower changes in user
distribution, traffic demand, and environmental conditions. Pinching
antennas therefore introduce a cross-layer control capability that links
radio geometry with higher-layer network operation.



\section{Pinching Antennas for EDMA} \label{sec: edma}

In this section, we investigate how pinching antennas can enable EDMA. We first introduce the basic concept of EDMA and its realization using pinching antennas. We then discuss the evolution toward artificial intelligence (AI)-native EDMA and its implications for access protocol and scheduling design.

\subsection{Introduction to EDMA}

Multiple-access techniques are a cornerstone of wireless communication systems because they determine how scarce radio resources are shared among users and fundamentally influence both transmission design and network operation. Conventional multiple-access schemes distinguish users through different resource domains. Time-division multiple access (TDMA) and frequency-division multiple access (FDMA) allocate orthogonal time or frequency resources, thereby suppressing multiple-access interference at the cost of limiting the resources available to each user. Code-division multiple access (CDMA), space-division multiple access (SDMA), and non-orthogonal multiple access (NOMA) allow users to share the same time-frequency resources by exploiting the code, spatial, and power domains, respectively, but generally require additional signal processing to separate the users. These developments motivate the exploration of additional design dimensions for supporting multiuser transmission.

EDMA provides a new perspective by exploiting differences in the propagation environments experienced by different users \cite{ding2026environment}. Wireless propagation conditions, including LoS availability, blockage, large-scale path loss, multipath propagation, and directional transmission characteristics, can create substantial differences between desired and interfering links, as illustrated in Fig. \ref{fig: pa three cases}(a). EDMA exploits these differences to strengthen user separation and suppress multiple-access interference. Users experiencing sufficiently isolated propagation conditions can therefore be served simultaneously over the same time-frequency resources. EDMA differs from conventional environment-aware resource allocation. Environment-aware schemes generally estimate the channel conditions created by a given environment and adapt transmission accordingly. In EDMA, the propagation environment plays a more direct role in determining how users are separated. Existing environmental features can be exploited when selecting users and transmission configurations, while flexible antennas and reconfigurable wireless infrastructure can further adjust how signals enter and propagate through the environment. Multiple-access requirements and propagation conditions can therefore be considered jointly rather than sequentially.

\begin{figure}[!t]
    \centering
    \includegraphics[width=0.82\linewidth]{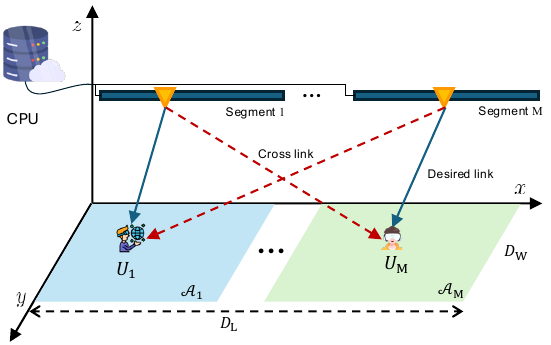}
    \captionsetup{justification=justified,singlelinecheck=false,font=small}
    \caption{Illustration of a pinching-antenna-enabled EDMA system, where
independently fed waveguide segments simultaneously serve users in
different service regions over the same time-frequency resource, while
environmental blockage weakens the cross-links and suppresses multiuser
interference.}
    \label{fig: edma model}
\end{figure}

Pinching antennas provide a particularly suitable platform for realizing EDMA because they can reconfigure radiation locations over extended spatial ranges. By selecting appropriate radiation locations, a pinching-antenna system can strengthen desired links while increasing the likelihood that interfering links are blocked or experience severe propagation loss, as illustrated in Fig. \ref{fig: edma model}. This allows multiple users to share the same radio resources with reduced interference while reducing the need for complex precoding or multiuser detection. Although EDMA is not limited to pinching antennas, their large-scale radiation-location reconfigurability provides a direct mechanism for creating the propagation differences needed for user separation. More broadly, EDMA changes the role of the propagation environment in multiple-access design. Rather than being treated solely as an external condition that the communication system must accommodate, the environment becomes an additional design dimension that can be sensed, exploited, and, when possible, reconfigured according to multiuser communication requirements \cite{ding2026environment2}. This perspective underpins the pinching-antenna-enabled EDMA realizations and extensions discussed in the remainder of this section.

\subsection{EDMA Realization Using Pinching Antennas}
\label{subsec:edma_realization}

The feasibility of realizing EDMA using pinching antennas was first
investigated in \cite{ding2026environment}. The basic idea is to
configure the pinching-antenna locations to maintain favorable desired links while
weakening cross-links by exploiting distance-dependent path loss and
LoS blockage, thereby enabling multiple users to reuse the same
time-frequency resource. To explain this realization, we next consider
a representative system model and use it to illustrate the key design
principles and performance insights.

\subsubsection{System Architecture and Channel Model}

Consider a rectangular service area with length $D_{\mathrm{L}}$ and
width $D_{\mathrm{W}}$, which is divided into $M$ non-overlapping
regions $\{\mathcal{A}_m\}_{m=1}^{M}$, as depicted in Fig. \ref{fig: edma model}. One user, denoted by $U_m$, is
scheduled in each region $\mathcal{A}_m$. A dielectric waveguide
deployed above the service area is accordingly divided into $M$
segments. Each segment has an independent feed and radio-frequency
chain, and one pinching antenna is activated on segment $m$ to serve $U_m$. The
users' signals can therefore be independently transmitted or processed
through different waveguide segments, although they occupy the same
wireless time-frequency resource. Let the locations of
user $U_k$ and pinching antenna $m$ be $ \boldsymbol{\psi}_k =
[x_k,y_k,0]^{\mathsf{T}} $ and $ \widetilde{\boldsymbol{\psi}}_m^{\mathrm{Pin}}
= [\widetilde{x}_m,0,d_v]^{\mathsf{T}}$
respectively. The distance between $U_k$ and pinching antenna $m$ is given by
\begin{align}
r_{k,m}
=
\|
\widetilde{\boldsymbol{\psi}}_m^{\mathrm{Pin}}
-
\boldsymbol{\psi}_k
\|
=
\sqrt{(x_k-\widetilde{x}_m)^2+y_k^2+d_v^2}.
\label{eq:edma_distance}
\end{align}
Let $\boldsymbol{\psi}_{0,m}$ denote the location of the feed point of
waveguide segment $m$. Following the channel model introduced in
Sec.~\ref{subsec: channel model}, the LoS channel between the feed
point of segment $m$ and $U_k$ is given by
\begin{align}
h_{k,m}^{\mathrm{LoS}}
=
\frac{\sqrt{\eta}}{r_{k,m}}
\exp\left[
-\mathrm{j}2\pi
\left(
\frac{r_{k,m}}{\lambda}
+
\frac{
\| \boldsymbol{\psi}_{0,m} - \widetilde{\boldsymbol{\psi}}_m^{\mathrm{Pin}} \|
}{\lambda_g}
\right)
\right].
\label{eq:edma_los_channel}
\end{align}
To account for random LoS blockage, the overall channel is expressed as
\begin{align}
h_{k,m}
=
\gamma_{k,m}h_{k,m}^{\mathrm{LoS}},
\label{eq:edma_channel}
\end{align}
where $\gamma_{k,m}\in\{0,1\}$ denotes the LoS indicator, with
\begin{align}
\Pr(\gamma_{k,m}=1)
=
\exp\left(-\beta r_{k,m}^2\right).
\label{eq:edma_los_probability}
\end{align}

\subsubsection{Realization of EDMA}

We next explain how the preceding architecture and channel
characteristics can support EDMA. For ease of illustration, we focus on
downlink transmission. Let $P_{\mathrm{T}}$ denote the total transmit
power and assume that it is equally allocated among the $M$ pinching antennas. The
pinching antennas simultaneously transmit independent signals to their associated
users. When multiuser interference is treated as noise, the
signal-to-interference-plus-noise ratio (SINR) at $U_k$ is given by
\begin{align}
\Gamma_k^{\mathrm{DL}} = \frac{\rho |h_{k,k}|^2}{1+\rho\sum_{m\neq k}|h_{k,m}|^2},
\label{eq:edma_dl_sinr}
\end{align}
where $\rho=P_{\mathrm{T}}/(M\sigma^2)$ denotes the per-pinching-antenna
transmit-to-noise ratio and $\sigma^2$ is the noise power. The
corresponding achievable rate is
\begin{align}
R_k^{\mathrm{DL}}=\log_2\left(1+\Gamma_k^{\mathrm{DL}}\right).
\label{eq:edma_dl_rate}
\end{align}

The SINR expression in \eqref{eq:edma_dl_sinr} shows that EDMA requires
a favorable difference between the desired and cross-links. To
illustrate how pinching-antenna configuration can create such a difference, consider
a simple user-aligned configuration  $\widetilde{x}_m=x_m,\ m=1,\ldots,M.$
Under this configuration, the desired-link distance between pinching antenna $k$ and
$U_k$ is
\begin{align}
r_{k,k} = \sqrt{y_k^2+d_v^2}.
\label{eq:edma_desired_distance}
\end{align}
For $m\neq k$, the cross-link distance between pinching antenna $m$ and $U_k$ is
\begin{align}
r_{k,m} = \sqrt{(x_k-x_m)^2+y_k^2+d_v^2}. \label{eq:edma_cross_distance}
\end{align}
Compared with the desired-link distance in
\eqref{eq:edma_desired_distance}, the cross-link distance in
\eqref{eq:edma_cross_distance} contains the additional horizontal
separation $(x_k-x_m)^2$. When users in different regions are
sufficiently separated, the cross-links therefore experience greater
path loss and lower LoS probabilities than the desired links. The
desired term in the numerator of \eqref{eq:edma_dl_sinr} can remain
strong, while the interference terms in the denominator are weakened
or blocked. Multiple users can consequently reuse the same
time-frequency resources without requiring coordinated multiuser
precoding.

To evaluate whether the simultaneous resource reuse enabled by EDMA
outperforms the resulting multiuser interference, we adopt pinching-antenna-assisted
TDMA as the main benchmark. Under
TDMA, the users are served in $M$ equal-duration time slots. Each user
therefore occupies $1/M$ of the time resource, while its associated pinching antenna
uses the total transmit power $P_{\mathrm{T}}$ during the allocated
slot. Assuming that the desired links remain LoS, the TDMA sum rate is given
by 
\begin{align}
R_{\mathrm{sum}}^{\mathrm{TDMA}} =
\frac{1}{M}
\sum_{m=1}^{M}
\log_2\left(
1+M\rho
\left|h_{m,m}^{\mathrm{LoS}}\right|^2
\right),
\label{eq:edma_tdma_rate}
\end{align}
where $\rho=P_{\mathrm{T}}/(M\sigma^2)$ denotes the per-user transmit SNR under equal power allocation in EDMA, and the total transmit power is $P_{\mathrm{T}}$ in both schemes. EDMA divides this power
equally among the $M$ pinching antennas and serves all users simultaneously, whereas
TDMA assigns the entire power to one pinching antenna at a time. Consequently, TDMA
avoids multiuser interference at the cost of the $1/M$ time-sharing
factor, while EDMA allows full-time resource reuse but experiences
interference among simultaneous transmissions. This tradeoff is
examined next.

\subsubsection{A Two-User Illustration}

To analytically illustrate this tradeoff, we consider a special case in
which two users are located underneath the waveguide, i.e.,
$y_1=y_2=0$, with each pinching antenna aligned with its associated user. Let
$\Delta=|x_2-x_1|$ denote their horizontal separation. The desired-link
distance is $d_v$, whereas the cross-link distance is
$\sqrt{\Delta^2+d_v^2}$. Assuming that the desired links remain LoS, the desired-link SNR and the cross-link SNR conditioned on LoS are respectively given by
\begin{align}
\Gamma_{\mathrm{d}}=\frac{\rho\eta}{d_v^2}, \ \
\Gamma_{\mathrm{c}}=\frac{\rho\eta}{\Delta^2+d_v^2},
\label{eq:edma_two_user_snrs}
\end{align}
and the LoS probability of each cross-link is
\begin{align}
p_{\mathrm{c}} = \exp\left[-\beta(\Delta^2+d_v^2)\right].
\label{eq:edma_cross_los_probability}
\end{align}

Assuming that the desired links are LoS, each user achieves the
interference-free rate $\log_2(1+\Gamma_{\mathrm{d}})$ when its
cross-link is blocked. When the cross-link is LoS, its achievable rate
becomes
\begin{align}
    \log_2\left(1+\frac{\Gamma_{\mathrm{d}}}{1+\Gamma_{\mathrm{c}}}\right).
\end{align}
The expected EDMA sum rate is therefore
\begin{align}
\overline{R}_{\mathrm{sum}}^{\mathrm{EDMA}}={}
&2(1-p_{\mathrm{c}})\log_2\left(1+\Gamma_{\mathrm{d}}\right)
\nonumber\\
&+2p_{\mathrm{c}}\log_2\left(1+\frac{\Gamma_{\mathrm{d}}}{1+\Gamma_{\mathrm{c}}}\right).
\label{eq:edma_two_user_sum_rate}
\end{align}
Under the same desired-link assumption, the corresponding TDMA sum rate
is given by
\begin{align}
R_{\mathrm{sum}}^{\mathrm{TDMA}}=\log_2\left(1+2\Gamma_{\mathrm{d}}\right).
\label{eq:edma_two_user_tdma_rate}
\end{align}

This example shows that user separation benefits EDMA through two
complementary effects. As $\Delta$ increases, the cross-link SNR
$\Gamma_{\mathrm{c}}$ decreases because of path loss, while its LoS
probability $p_{\mathrm{c}}$ decreases because of blockage. When
$p_{\mathrm{c}}\rightarrow0$, the EDMA sum rate approaches
\begin{align}\overline{R}_{\mathrm{sum}}^{\mathrm{EDMA}}\rightarrow2\log_2\left(1+\Gamma_{\mathrm{d}}\right),
\end{align}
which is greater than
$\log_2(1+2\Gamma_{\mathrm{d}})$ for $\Gamma_{\mathrm{d}}>0$. EDMA is therefore advantageous when the
environment provides sufficient cross-link isolation. In contrast, only when
the users are closely located and their cross-links are likely to remain
LoS, EDMA becomes interference-limited, and TDMA may achieve a higher sum
rate.

More generally, \cite{ding2026environment} investigated the two-user
case with users randomly located in their respective service regions.
Lower bounds were derived for the ergodic sum-rate gain of EDMA over
TDMA and the probability that EDMA achieves a higher instantaneous sum
rate than TDMA. Asymptotic analysis further characterized how the
performance gain depends on the service-area size and the likelihood of
LoS blockage. The results show that EDMA becomes more advantageous as
the service area increases or the cross-links become more likely to be
blocked.

\subsubsection{Pinching-Antenna Location Optimization and Performance Evaluations}

The user-aligned configuration, i.e., $\widetilde{x}_m=x_m, \ m \!=\!1,\ldots,M$, is effective when users are sufficiently
separated. When users in neighboring regions are closely located,
however, placing each pinching antenna directly above its associated user may create
strong cross-links. Optimizing the pinching-antenna locations can provide a better
balance between desired-link enhancement and interference suppression.
We next investigate how pinching-antenna location optimization can further improve
the downlink performance of EDMA. 

Let $\widetilde{\boldsymbol{x}}
=[\widetilde{x}_1,\ldots,\widetilde{x}_M]^{\mathsf T}$
collect the pinching-antenna locations, and define the average downlink rate of
$U_m$ as
\begin{align}
\overline{R}_m^{\mathrm{DL}}
(\widetilde{\boldsymbol{x}})
=
\mathbb{E}_{\boldsymbol{\gamma}}
\left[
R_m^{\mathrm{DL}}
(\widetilde{\boldsymbol{x}})
\right],
\label{eq:edma_average_dl_rate}
\end{align}
where the expectation is taken over the random blockage states. To
provide balanced service among users, the pinching-antenna locations are optimized
to maximize the minimum average downlink rate as
\begin{align}
\underset{\widetilde{\boldsymbol{x}},\,t}
{\max}\quad& t \label{eq:edma_dl_maxmin_optimization}\\
\st\quad& \overline{R}_m^{\mathrm{DL}}(\widetilde{\boldsymbol{x}})\geq t,
\quad m=1,\ldots,M,\nonumber\\
&\widetilde{x}_m\in\mathcal{X}_m,\quad m=1,\ldots,M,\nonumber
\end{align}
where $\mathcal{X}_m$ denotes the feasible activation interval on
waveguide segment $m$.

Problem~\eqref{eq:edma_dl_maxmin_optimization} is nonconvex because
each pinching-antenna location affects both its desired link and the interference
received by other users. Therefore, all pinching-antenna locations are coupled and
must be jointly optimized. This difficulty was addressed in \cite{ding2026environment} by first adopting a tractable lower-bound approximation of the average rates and then applying successive convex approximation (SCA). At each iteration, the nonconvex rate constraints
are replaced by convex surrogate constraints constructed around the
current pinching-antenna locations. The resulting convex problem is solved to update
all pinching-antenna locations, and the procedure continues until convergence.
Numerical comparisons with exhaustive search show that the SCA method
achieves near-optimal performance for the considered settings.

\begin{figure}[!t]
    \centering
    \includegraphics[width=0.82\linewidth]{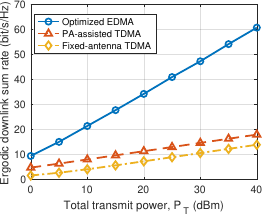}
    \captionsetup{justification=justified,singlelinecheck=false,font=small}
    \caption{Ergodic downlink sum rate versus the total transmit power for
optimized EDMA, pinching-antenna-assisted TDMA, and conventional fixed-antenna TDMA, where PA denotes pinching antenna.}
    \label{fig:edma_performance}
\end{figure}

Fig.~\ref{fig:edma_performance} illustrates the downlink performance
of EDMA in a service region with dimensions $D_L=100$~m and
$D_W=20$~m. The region is divided into $M=4$ equal service regions
along the $x$-axis, with one user uniformly distributed in each region.
The carrier frequency is $28$~GHz, the waveguide height is $d_v=3$~m,
the blockage parameter is $\beta=0.02$, and the noise power is
$-90$~dBm. The results are averaged over $100$ independent user-location
realizations. The figure compares optimized EDMA, pinching-antenna-assisted TDMA,
and conventional fixed-antenna TDMA. Optimized EDMA jointly determines
the pinching-antenna locations using the optimization method described above.
Pinching-antenna-assisted TDMA activates a pinching antenna horizontally aligned with each
scheduled user and serves the users in orthogonal time slots, whereas
conventional TDMA employs a fixed antenna at the center of the service
region.

As shown in Fig.~\ref{fig:edma_performance}, optimized EDMA achieves
the highest ergodic downlink sum rate over the considered transmit-power
range. By jointly adjusting the pinching-antenna locations, EDMA maintains strong
desired links while reducing the strength and LoS availability of
cross-links. This allows the four users to be served simultaneously
with controlled multiuser interference, providing a clear sum-rate
gain over pinching-antenna-assisted TDMA.
Pinching-antenna-assisted TDMA also outperforms conventional fixed-antenna TDMA
because its radiation location can be horizontally aligned with each
scheduled user, thereby reducing the desired-link propagation distance.
The additional gain of optimized EDMA over pinching-antenna-assisted TDMA therefore
mainly results from simultaneous resource reuse and pinching-antenna-based
interference suppression. The increasing performance advantage at
higher transmit power further shows that effective cross-link isolation
is essential for preventing multiuser interference from limiting the
benefit of simultaneous transmission. Overall, the results demonstrate
that pinching-antenna location optimization enables EDMA to exploit environmental
separation while preserving strong desired links.

Beyond this representative example, \cite{ding2026environment}
provided a broader set of analytical and numerical results for EDMA.
The effects of the service-area size, LoS blockage parameter, number of
users, user distribution, and transmit power were investigated under
both regularly separated and clustered-user scenarios. The results
showed that a larger service area and a higher likelihood of cross-link
blockage generally improved the gain of EDMA, whereas serving too many
closely located users could introduce severe multiuser interference.
The developed pinching-antenna location optimization methods were also compared with
exhaustive search and were shown to achieve near-optimal performance in
the considered settings. 
Moreover, EDMA can also be applied to uplink transmission, where
multiple users transmit simultaneously to pinching antennas activated on their
respective waveguide segments, and the pinching-antenna locations can be optimized to
strengthen the desired links while suppressing multiple-access
interference.

\subsubsection{Extension of EDMA to NLoS Channels}

The preceding discussion assumes that a blocked cross-link contributes
negligible interference. In practice, however, blockage may replace the
LoS component with residual NLoS scattering rather than completely
eliminating the link \cite{xu2026_losnlos,wang2025pinching-nlos}. Moreover, desired links may also become NLoS,
introducing random rate degradation and communication outages. EDMA
must therefore account for both the interference-suppression benefit of
LoS blockage and the residual interference and reliability loss caused
by NLoS propagation.

To address this issue, \cite{mao2026outage} investigated an
outage-constrained EDMA design accounting for probabilistic LoS
blockage and random NLoS scattering. The pinching-antenna locations and transmit
powers were jointly optimized to maximize the total information rate
subject to the per-user outage constraints
$\Pr\{R_m<R_m^{\mathrm{tar}}\}\leq\epsilon_m$, where
$R_m^{\mathrm{tar}}$ and $\epsilon_m$ denote the target rate and the
maximum allowable outage probability of $U_m$, respectively. For the
two-user case, closed-form outage probabilities were derived and a
projected-gradient method was developed. For the general multiuser
case, tractable outage approximations were obtained and an SCA-based
method was proposed. Numerical results showed that both methods
achieved near-optimal performance compared with exhaustive search and
that EDMA maintained a clear advantage over pinching-antenna-assisted TDMA and
conventional fixed-antenna TDMA in the presence of random LoS blockage
and NLoS scattering. These results demonstrate that EDMA can still
exploit environmental blockage when blocked links retain residual NLoS
components, provided that pinching-antenna configuration and resource allocation
explicitly account for the resulting channel uncertainty.

\subsection{Toward AI-Native EDMA}

Practical EDMA operation requires knowledge of the propagation
environment and the joint configuration of flexible antennas and
communication resources. These tasks become particularly challenging
in dynamic multiuser networks, where user locations, blockage
conditions, traffic demand, and service requirements vary over time.
AI can support the evolution of EDMA in two stages \cite{ding2026environment2}. In AI-assisted
EDMA, AI tools are used to address specific sensing and optimization
tasks within a predefined EDMA framework. In AI-native EDMA, AI is
instead embedded into the multiple-access architecture to enable
continuous interaction between the network and its propagation
environment.

\subsubsection{AI-Assisted EDMA}
One important application of AI is environment sensing and
understanding. Conventional channel estimation provides end-to-end
channel coefficients but generally does not reveal the environmental
factors that produce them. EDMA requires more explicit knowledge,
including user locations, blockage conditions, propagation paths,
environmental objects, and mobility patterns. Measurements obtained
from channel estimation, integrated sensing and communication, channel
knowledge maps, digital twins, and other sensing mechanisms can be
processed by AI to extract such propagation-relevant knowledge. For
example, AI can identify which objects block particular cross-links and
which radiation locations can establish strong desired links.

AI can also assist EDMA resource allocation. The joint design of user
grouping, scheduling, pinching-antenna activation and positioning, beamforming, and
power allocation generally involves highly nonconvex problems with
mixed discrete and continuous variables. Moreover, these variables may
belong to different protocol layers and operate over different time
scales. Learning-based methods can complement model-based optimization
by identifying promising user groups, learning good initial solutions,
or directly producing high-quality decisions when repeated numerical
optimization is too costly.

\subsubsection{AI-Native EDMA}
AI-native EDMA goes beyond applying separate AI tools to individual
tasks. It treats AI as an integral part of the multiple-access
architecture and establishes a closed loop consisting of sensing,
learning, prediction, and reconfiguration. The network first senses
users and relevant environmental features from communication and
sensing measurements. It then learns how these features affect desired
and interfering links and how users can be separated through the
environment. Based on historical and current observations, it predicts
future user locations, traffic demand, mobility, and propagation
conditions. Finally, the predicted knowledge is used to reconfigure pinching antenna
locations, user scheduling, power allocation, and other communication
resources.

This process is continuous rather than a one-time operation. The
effects of each reconfiguration are sensed again and used to update the
learned environment model and subsequent decisions. AI-native EDMA
therefore transforms the propagation environment from a passively
observed channel condition into a multiple-access resource that is
continuously sensed, learned, predicted, and reconfigured according to
network demand.

\subsection{EDMA-Assisted Access Protocol and Scheduling}
\label{subsec:edma_access_scheduling}

The preceding discussions mainly consider EDMA in scheduled
multiuser transmission, where the active users are known
before pinching-antenna configuration and resource allocation. Practical
wireless networks must also support sporadically active users
and time-varying service demand. In these cases, EDMA cannot
be applied only after the served users have been determined.
Its environment-configuration dimension must instead be
incorporated into the access and scheduling procedures.

From this perspective, EDMA provides protocol operation with
an additional configurable resource. During random access,
different pinching antenna configurations create environment-dependent
access opportunities that can help separate uncoordinated
users. During dynamic scheduling, different pinching-antenna activation
patterns provide spatial service actions that can be selected
according to evolving queue states. In this subsection, we discuss these two
representative forms of EDMA-assisted protocol operation.

\subsubsection{Configuration-Aware Random Access}

Conventional random-access protocols allow sporadically
active users to contend over time slots, frequency resources,
or preambles while the underlying propagation environment
remains largely fixed. EDMA-assisted random access also
exploits the pinching-antenna configuration domain during contention. A
representative example is the segmented waveguide-enabled
pinching-antenna system considered in \cite{shan2026access},
where a long waveguide is divided into independently
controlled segments and one pinching antenna can be activated on each
segment.

Let $\mathcal M$ denote the segment set and $\mathcal P_m$
the candidate pinching antenna locations on segment $m$. A segment--pinching antenna
pair $(m,p)$ defines an elementary configuration, and all such
pairs form the configuration space
$\mathcal C=\{(m,p):m\in\mathcal M,\ p\in\mathcal P_m\}$.
Different configurations induce different channel responses
for the same user, while the same configuration can affect
different users differently. From an EDMA perspective, this
configuration-dependent channel diversity creates a family of
environment-domain access opportunities through which users
can be distinguished during contention.

Exhaustively probing the entire configuration space would,
however, introduce considerable pilot and switching overhead.
The oracle-assisted framework in \cite{shan2026access}
therefore probes only a sparse subset of pinching antenna configurations and
uses the resulting observations to infer the remaining
configuration-dependent channels. During random access, each
active user can then identify favorable configuration
opportunities rather than contending blindly. The channel
oracle therefore provides the environment knowledge required
to exploit EDMA before conventional user scheduling has been
completed.

Two EDMA-assisted access mechanisms are considered according
to the available RF hardware. In segment aggregation, several
segments are connected to a single RF chain through analog
aggregation. Randomized pinching-antenna activation across access attempts
creates configuration diversity with low RF complexity.
However, signals received through different segments remain
coupled in the same analog observation, which limits
collision-resolution capability.

In segment multiplexing, the active segments are connected to
dedicated RF chains. A deterministic access codebook
specifies the pinching-antenna configuration used in each access slot, and
each user employs its channel oracle to select a favorable
slot. Because the observations from different segments are
preserved separately, the receiver can better distinguish and
resolve simultaneous access attempts, at the cost of
additional RF chains and baseband processing.

These mechanisms illustrate how EDMA can assist random
access through different architecture-dependent tradeoffs.
Segment aggregation exploits randomized environment
configurations to provide low-complexity access, whereas
segment multiplexing uses deterministic configuration
codewords and additional RF dimensions to improve access
reliability and collision resolution. Sparse channel-oracle
acquisition further balances the available environment
knowledge against pilot and switching overhead. The results
in \cite{shan2026access} show that reliable configuration
prediction can be achieved with substantially reduced
training, while the preferred access architecture depends on
the offered load and RF-complexity budget.

\subsubsection{Queue-Aware Online Pinching Antenna Scheduling}

EDMA can also assist online scheduling by treating feasible pinching antenna
activation patterns as spatial service actions. This is
particularly relevant to vehicular and dense-access networks,
where service demand can shift rapidly among different
regions. The main challenge is that pinching-antenna activation is subject
to discrete spatial constraints, such as allowing only one
active pinching antenna on each waveguide, while the arrival and service
statistics may not be known in advance.

The schedule-as-you-learn strategy developed in
\cite{yu2026online} learns the required service rates from
observed arrivals and previous scheduling decisions. These
requirements are mapped to feasible pinching-antenna activation patterns
and translated into slot-by-slot scheduling actions. EDMA
therefore allows the scheduler to adapt not only conventional
time-frequency resources but also the spatial propagation
configuration according to the queue state. The results show
that this approach can maintain bounded queue backlogs and
provide more balanced delay performance than random
activation and a MaxWeight-based benchmark.

\subsubsection{Protocol-Level Interpretation}

The preceding examples extend EDMA beyond scheduled
physical-layer transmission. In random access, the EDMA
configuration space provides environment-dependent access
opportunities through which sporadically active users can
contend and be distinguished. In online scheduling, the same
configuration space provides spatial service actions that can
be selected according to evolving queue states.

Conventional access protocols and schedulers mainly allocate
time-frequency resources over a fixed propagation topology.
EDMA-assisted protocols additionally allow the propagation
configuration itself to participate in contention resolution
and service scheduling. This creates a cross-layer design
space in which channel acquisition, access contention, RF
complexity, queue states, and pinching-antenna activation are jointly
coupled.

The resulting tradeoffs depend strongly on the pinching antenna
architecture. Sparse channel-oracle acquisition balances
configuration knowledge against training overhead. Segment
aggregation and segment multiplexing trade RF complexity for
access reliability and collision resolution. Over a longer time
scale, online pinching antenna scheduling balances adaptation speed, queue
stability, and delay under discrete activation constraints.
Overall, these examples show that EDMA configurations can
serve not only as physical-layer optimization variables but
also as protocol-level resources for random access and
queue-aware scheduling.

\subsection{Summary and Outlook}

EDMA exploits radiation-location reconfigurability and spatially
varying propagation conditions to create favorable differences
between desired and cross-links. By positioning pinching antennas close to their
associated users or at locations with favorable LoS connectivity, the
desired links can remain strong, while the cross-links are weakened
through greater propagation distances, environmental blockage, or
directional radiation. Multiple users can consequently reuse the same
time-frequency resources with reduced interference. The analytical
and numerical examples presented in this section further show that the
resulting gain depends on user separation, blockage conditions,
transmit power, and pinching-antenna configuration. Beyond the basic downlink LoS
setting, EDMA can also support uplink transmission and remain
effective under random LoS/NLoS channels. AI-assisted environment
sensing, prediction, and resource allocation can further support EDMA
when propagation conditions, user locations, and traffic demand vary
over time. These developments broaden EDMA from a static
interference-management mechanism toward an environment-aware
multiple-access framework.

Several challenges remain before EDMA can be incorporated into
large-scale wireless networks. Future research should investigate
scalable user grouping, region partitioning, pinching-antenna configuration, power
allocation, and interference management for multi-antenna and
multi-waveguide systems. Robust designs are also needed under
imperfect environment knowledge, user mobility, time-varying
blockage, and practical pinching-antenna reconfiguration constraints. At the
protocol level, EDMA should be integrated with user discovery, random
access, scheduling, channel acquisition, and retransmission
mechanisms, while its interaction with established multiple-access
techniques such as NOMA and SDMA requires further study. Developing
such cross-layer methods will be important for translating
environment-based user separation into reliable and scalable network
operation.

\section{Pinching-Antenna-Enabled Multi-Cell Transmission}\label{sec: multicell}
This section extends the discussion from multiuser transmission within a cell to pinching-antenna-enabled operation across multiple cells. In a multi-cell network, pinching-antenna configuration affects not only the desired links within individual cells but also inter-cell interference, serving relationships, and the effective geometry of the cellular network. Radiation-location reconfigurability can therefore enable new forms of power-efficient transmission, cell reconfiguration, traffic offloading, and network-topology adaptation. We first motivate the use of pinching antennas in multi-cell networks and then present three representative examples, including geometry-reconfigurable low-power transmission, cell reconfiguration and traffic offloading, and pinching-antenna-enabled large-scale cell-boundary reconfiguration. Finally, we summarize the main insights and discuss the research opportunities for pinching-antenna-enabled multi-cell networks.

\subsection{Motivation for Multi-Cell Pinching-Antenna Systems}
Multi-cell networks require the coordination of desired-link transmission, inter-cell interference, user association, and traffic distribution across different cells. Existing architectures, including heterogeneous, cloud, and fog radio access networks, improve multi-cell operation through dense access-point deployment, centralized resource pooling, coordinated processing, and edge-based control. These
mechanisms provide substantial flexibility in determining which transmission points and radio resources serve each user. However, the physical locations of these transmission points generally remain fixed after network deployment.

The radiation-location reconfigurability of pinching antennas introduces an additional degree of freedom for multi-cell design. Adjusting the pinching-antenna location in one cell can simultaneously modify its desired link and the interference links toward users in neighboring cells. It can also change the effective service range of a BS, allowing users to be served locally, jointly assisted by multiple BSs, or offloaded to neighboring cells according to traffic and propagation conditions. When pinching antennas are deployed across many cells, their coordinated configuration can further reshape cell boundaries and the large-scale spatial structure of the network.

Pinching-antenna-enabled multi-cell design therefore extends conventional resource allocation and user-association optimization over a fixed radio topology to the joint configuration of radio resources, serving relationships, and radiation geometry. Realizing this potential requires pinching-antenna locations to be coordinated with transmit power, user association, traffic offloading, and inter-cell cooperation, subject to the available waveguide deployments and coordination capabilities.

\subsection{Geometry-Reconfigurable Low-Power Transmission}
Conventional multi-cell transmission is fundamentally constrained by the geometry between fixed radio sites and their associated users. In particular, a cell-edge user may be located far from its serving BS and therefore require high transmit power to satisfy a prescribed QoS requirement. Increasing the transmit power strengthens the desired link but simultaneously increases the interference experienced by users in neighboring cells. Conventional interference-management architectures alleviate this problem through power control, coordinated transmission, data sharing, or centralized processing, but the physical transmission locations generally remain fixed. Pinching antennas provide an additional mechanism by allowing each cell to reconfigure its radiation location according to the spatial distribution of users. In this subsection, we show how pinching-antenna-enabled geometry reconfiguration can reduce the network-wide transmit power required for QoS-guaranteed multi-cell transmission.

\subsubsection{Multi-Cell Transmission Model}
Consider a downlink network consisting of $M$ cells, with one user $U_m$ served in cell $m$. Each BS feeds a waveguide deployed within its cell, and one pinching antenna is activated at a configurable location on that waveguide. The BS in cell $m$ transmits only the signal intended for $U_m$, while its transmission causes inter-cell interference to users in the other cells. The cells may coordinate their pinching-antenna locations and transmit powers, but they do not jointly transmit user data. Therefore, data sharing and phase-coherent synchronization among the BSs are not required.

Let the location of $U_m$ be $\boldsymbol{\psi}_m=[x_m,y_m,0]^{\mathsf T}$.  The pinching antenna activated on waveguide $n$ is located at  $\widetilde{\boldsymbol{\psi}}_n^{\rm Pin} = [\widetilde{x}_n,\widetilde{y}_n,d_v]^{\mathsf T}$,  where $\widetilde{y}_n$ and $d_v$ are determined by the deployment of waveguide $n$, whereas $\widetilde{x}_n$ can be adjusted within its feasible waveguide segment. Let $\boldsymbol{\psi}_{0,n}$ denote the feed point of waveguide $n$. Following the channel model in Sec.~\ref{subsec: channel model}, the LoS channel from the pinching antenna in cell $n$ to $U_m$ is given by
\begin{align}
h_{m,n}=\frac{\sqrt{\eta} \exp\Big[-\mathrm{j}2\pi\Big(\frac{\|\boldsymbol{\psi}_m-
\widetilde{\boldsymbol{\psi}}_n^{\rm Pin}\|}{\lambda}
+\frac{\|\boldsymbol{\psi}_{0,n}-\widetilde{\boldsymbol{\psi}}_n^{\rm Pin}\|}{\lambda_g}\Big)\Big]}{\|\boldsymbol{\psi}_m-\widetilde{\boldsymbol{\psi}}_n^{\rm Pin}\|}.
\label{eq:multicell_channel}
\end{align}

Let $s_n$ denote the unit-power signal transmitted in cell $n$, i.e., $\mathbb{E}[|s_n|^2]=1$, and let $P_n$ denote the corresponding transmit power. The received signal at $U_m$ is given by
\begin{align}
y_m = \sqrt{P_m}h_{m,m}s_m + \sum_{n\neq m}\sqrt{P_n}h_{m,n}s_n + w_m,
\label{eq:multicell_received_signal}
\end{align}
where $w_m$ is additive noise with variance $\sigma^2$. When inter-cell interference is treated as noise, the achievable rate of $U_m$ is
\begin{align}
R_m = \log_2\left(1 + \frac{P_m|h_{m,m}|^2}{\sum_{n\neq m}P_n|h_{m,n}|^2+\sigma^2}\right).
\label{eq:multicell_rate}
\end{align}

Equations~\eqref{eq:multicell_channel}--\eqref{eq:multicell_rate} highlight the additional design freedom provided by pinching antennas. Moving the pinching antenna in cell $m$ changes both its desired-link gain $|h_{m,m}|^2$ and the interference-link gains $\{|h_{n,m}|^2\}_{n\neq m}$ toward users in the other cells. The pinching-antenna locations can therefore be coordinated across cells to strengthen the desired links while limiting inter-cell interference, thereby reducing the transmit power required to satisfy the users' QoS requirements.

\subsubsection{Problem Formulation and Optimization}
A representative design objective is to minimize the total transmit power while satisfying the QoS requirement of every user. Let $R_{\rm t}$ denote the common target rate. The joint pinching-antenna location and power optimization problem is formulated as
\begin{subequations}
\begin{align}
\underset{\mathbf{p},\widetilde{\mathbf{x}}}{\min}\quad
& \sum_{m=1}^{M}P_m
\label{eq:multicell_power_obj}\\
\st\quad
& R_m\geq R_{\rm t},\quad m=1,\ldots,M,
\label{eq:multicell_rate_con}\\
& \widetilde{x}_{m,\min}\leq\widetilde{x}_m
\leq\widetilde{x}_{m,\max},\quad m=1,\ldots,M,
\label{eq:multicell_position_con}\\
& \mathbf{p}\succeq\mathbf{0},
\label{eq:multicell_power_con}
\end{align}
\end{subequations}
where $\mathbf{p}=[P_1,\ldots,P_M]^{\mathsf T}$ and $\widetilde{\mathbf{x}} =[\widetilde{x}_1,\ldots,\widetilde{x}_M]^{\mathsf T}$. The problem jointly determines how strongly each cell transmits and where its signal is radiated.

For fixed pinching-antenna locations, the optimal power allocation can be obtained explicitly. Define $\epsilon=2^{R_{\rm t}}-1$ and the normalized interference-coupling matrix $\mathbf{G}\in\mathbb{R}^{M\times M}$, whose entries are
\begin{align}
[\mathbf{G}]_{m,n} = \begin{cases}
\dfrac{|h_{m,n}|^2}{|h_{m,m}|^2}, & n\neq m,\\[2mm]
0, & n=m.
\end{cases}
\label{eq:multicell_G}
\end{align}
Further define
\begin{align}
\mathbf{a} = \epsilon\sigma^2 \left[\frac{1}{|h_{1,1}|^2},\ldots,\frac{1}{|h_{M,M}|^2}\right]^{\mathsf T}.
\label{eq:multicell_a}
\end{align}
At the minimum-power solution, all rate constraints are active, and the power vector satisfies
\begin{align}
\left(\mathbf{I}_M-\epsilon\mathbf{G}\right)\mathbf{p} = \mathbf{a}.
\label{eq:multicell_power_linear}
\end{align}
When the target rates are feasible, the optimal power allocation for a given pinching-antenna configuration is therefore
\begin{align}
\mathbf{p}^{\star}=\left(\mathbf{I}_M-\epsilon\mathbf{G}\right)^{-1}\mathbf{a}.
\label{eq:multicell_power_closed_form}
\end{align}
The pinching-antenna locations determine both $\mathbf{a}$, through the desired-link gains, and $\mathbf{G}$, through the relative interference-link gains. Substituting~\eqref{eq:multicell_power_closed_form} into~\eqref{eq:multicell_power_obj} thus reduces the joint design to an optimization over the pinching-antenna locations. The resulting problem remains nonconvex because the channel gains depend nonlinearly on these locations. The cross-entropy method adopted in~\cite{ding2026toward} addresses this problem by sampling candidate pinching-antenna configurations, evaluating their required transmit power using~\eqref{eq:multicell_power_closed_form}, and iteratively updating the sampling distribution according to the best-performing candidates. The sampling distribution can be initialized around the user locations, thereby incorporating useful geometric prior information.

Further insight can be obtained from the two-cell case. For $M=2$, the target rate is feasible only if
\begin{align}
\frac{|h_{1,1}|^2|h_{2,2}|^2} {|h_{1,2}|^2|h_{2,1}|^2} > \epsilon^2.
\label{eq:multicell_feasibility}
\end{align}
Define the desired-to-cross-link ratios associated with the two pinching antennas as
\begin{align}
g_1 &= \frac{|h_{1,1}|^2}{|h_{2,1}|^2} = \frac{\|\boldsymbol{\psi}_2- \widetilde{\boldsymbol{\psi}}_1^{\rm Pin}\|^2} {\|\boldsymbol{\psi}_1- \widetilde{\boldsymbol{\psi}}_1^{\rm Pin}\|^2},
\label{eq:multicell_g1}\\
g_2 &= \frac{|h_{2,2}|^2}{|h_{1,2}|^2} =
\frac{\|\boldsymbol{\psi}_1- \widetilde{\boldsymbol{\psi}}_2^{\rm Pin}\|^2}
{\|\boldsymbol{\psi}_2- \widetilde{\boldsymbol{\psi}}_2^{\rm Pin}\|^2}.
\label{eq:multicell_g2}
\end{align}
The feasibility condition can thus be written as $g_1g_2>\epsilon^2$. Hence, each pinching antenna should be positioned close to its intended user while remaining sufficiently far from the user in the neighboring cell. Moreover, $g_1$ and $g_2$ depend separately on the locations of pinching antennas $1$ and $2$, respectively. This motivates a low-complexity geometry-based design in which each pinching antenna maximizes its own desired-to-cross-link ratio. For a one-dimensional waveguide, the solution can be obtained by comparing the feasible stationary points and waveguide boundaries \cite{ding2026toward}. Although this design is generally suboptimal for total-power minimization, it captures the main geometric tradeoff and provides a low-complexity alternative to joint cross-entropy optimization.
\begin{figure}[!t]
    \centering
    \includegraphics[width=0.82\linewidth]
    {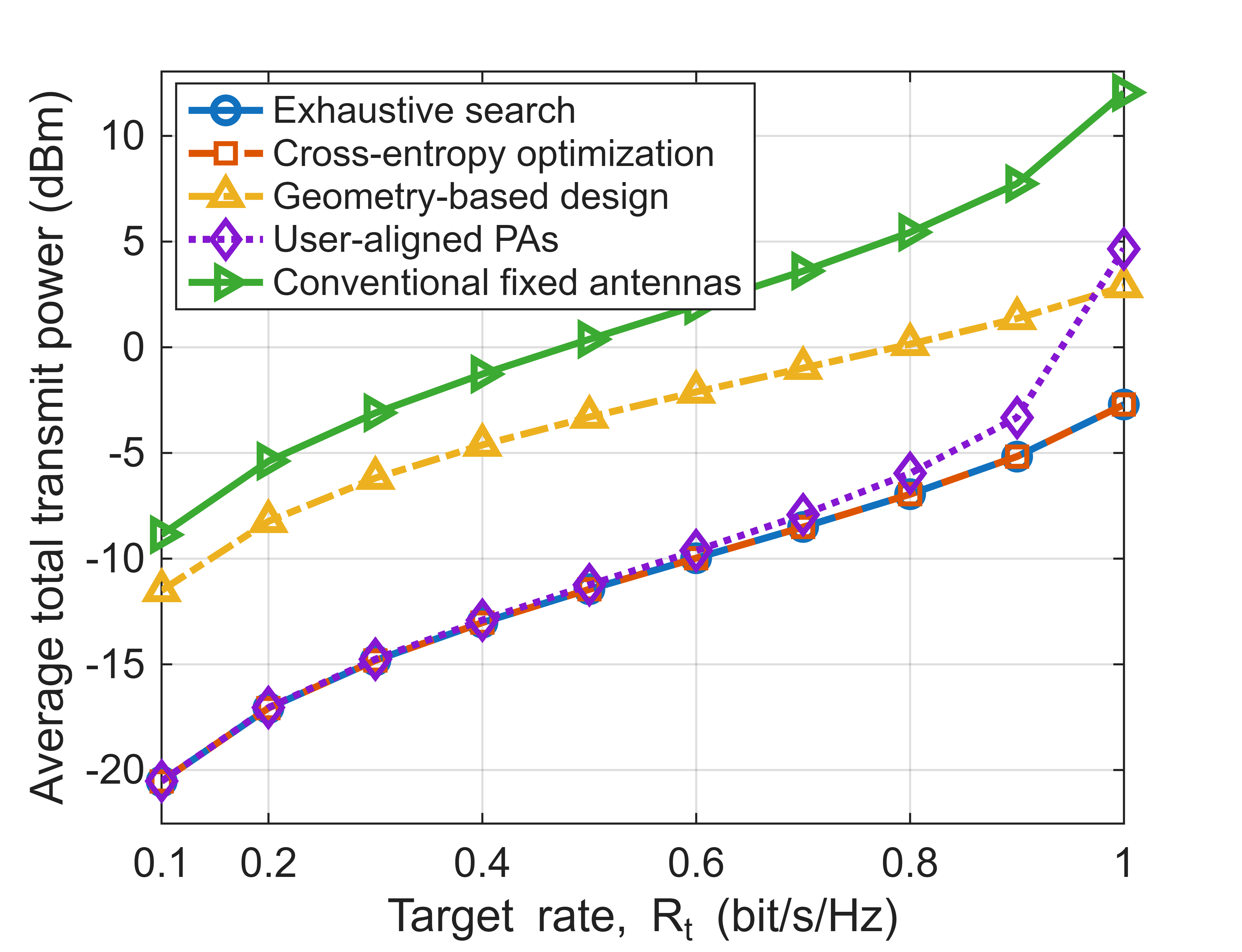}
    \captionsetup{justification=justified,
    singlelinecheck=false,font=small}
    \caption{Average total transmit power versus the target rate for pinching-antenna-enabled and conventional multi-cell transmission.}
    \label{fig:multicell_low_power}
\end{figure}

\subsubsection{Numerical Illustration and Design Insights}
We next consider a two-cell setting with clustered users to illustrate the preceding analytical insights. The service region has $D_{\rm W}=20$~m and $D_{\rm L}=100$~m, and the two cells occupy its left and right halves. The users are located near the common cell boundary, with $x_1\sim\mathcal{U}(-2,0)$ and $x_2\sim\mathcal{U}(0,2)$, while their remaining coordinates are uniformly generated within their respective cells. The noise power is $-70$~dBm, and the results are averaged over 100 independent user realizations. For the cross-entropy method, $N_{\rm CE}=1000$ candidate configurations are generated in each iteration, the best $N_{\rm best}=10$ configurations are retained, and the maximum number of iterations is $N_{\max}=10$.

The cross-entropy method is compared with exhaustive pinching-antenna location search, the low-complexity geometry-based design, a user-aligned scheme with $\widetilde{x}_m=x_m$, and a conventional scheme with one fixed antenna at the center of each cell. Fig.~\ref{fig:multicell_low_power} shows the average total transmit power versus the common target rate. The cross-entropy method essentially overlaps with exhaustive search over the entire considered range, demonstrating that it can effectively identify favorable pinching-antenna configurations with much lower search complexity.

At low and moderate target rates, the user-aligned scheme also performs close to the optimized schemes because shortening the desired links provides the dominant power-saving gain. As the target rate approaches $1$~bit/s/Hz, however, its required power increases sharply, indicating that positioning each pinching antenna directly above its user does not sufficiently account for the resulting cross-links. The geometry-based design requires more power at low and moderate rates because maximizing the desired-to-cross-link ratios does not directly minimize the total transmit power. Nevertheless, it consistently outperforms the conventional fixed-antenna scheme and becomes more effective than the user-aligned scheme at the highest considered target rate. This confirms that interference-aware pinching-antenna positioning becomes increasingly important as the QoS requirement becomes more stringent.

Overall, the results reveal two complementary roles of pinching-antenna positioning. Desired-link shortening provides most of the power-saving gain under moderate QoS requirements, whereas coordinated control of the cross-links becomes essential under stringent requirements. By balancing these two effects, optimized pinching-antenna positioning enables multi-cell transmission to move from a high-power ``shouting'' strategy toward low-power ``whispering.'' In this example, the serving relationships remain fixed; the next subsection considers cell reconfiguration and traffic offloading, where pinching antenna flexibility can also change which BSs serve the users.

\subsection{Cell Reconfiguration and Traffic Offloading}
The preceding subsection considered multi-cell transmission with fixed serving relationships. In practice, spatially uneven traffic may overload some cells while leaving neighboring cells with unused resources. Traffic offloading addresses this imbalance by transferring users from overloaded cells to underutilized ones. The radiation-location reconfigurability of pinching antennas provides further flexibility for this process. An underutilized cell can move its radiation point toward users in a neighboring cell, thereby extending its effective service region without relocating its BS, as illustrated in Fig. \ref{fig: pa three cases}(b). This subsection investigates the joint design of pinching-antenna configuration, traffic offloading, and resource allocation.

\subsubsection{System and Offloading Models}
To illustrate the main idea, we consider a two-cell downlink network. BS~$0$ is underutilized and connected to a waveguide on which one pinching antenna can be activated, whereas BS~$1$ is overloaded and employs a conventional fixed antenna. User $U_0$ is originally served by BS~$0$, while user $U_1$, originally associated with BS~$1$, is considered for offloading to BS~$0$. By moving the pinching antenna toward $U_1$, BS~$0$ can extend its effective service region into Cell~$1$ and potentially serve $U_1$ with lower transmit power.

Let the locations of $U_i$, $i\in\{0,1\}$, and BS~$1$ be denoted by $\boldsymbol{\psi}_i=[x_i,y_i,0]^{\mathsf T}$ and $\boldsymbol{\psi}_1^{\rm BS}$, respectively. The pinching-antenna location is expressed as $\widetilde{\boldsymbol{\psi}}^{\rm Pin} = [\widetilde{x},\widetilde{y},d_v]^{\mathsf T}$, where $\widetilde{y}$ is determined by the waveguide deployment and $\widetilde{x}$ can be adjusted within the feasible interval $\mathcal{X}=[\widetilde{x}_{\min},\widetilde{x}_{\max}]$.

Without traffic offloading, $U_1$ is served only by BS~$1$. Its achievable rate is given by
\begin{align}
R_1^{\rm No\text{-}Off} = \log_2\left(1+\frac{\eta P_1^{\rm No\text{-}Off}}{\sigma^2\left\|\boldsymbol{\psi}_1-\boldsymbol{\psi}_1^{\rm BS}\right\|^2}\right).
\label{eq:offloading_no_off_rate}
\end{align}
For a target rate $R_{\rm t}$, the minimum required power is given by
\begin{align}
P_1^{\rm No\text{-}Off}=\xi\left\|\boldsymbol{\psi}_1-\boldsymbol{\psi}_1^{\rm BS}\right\|^2,
\ \
\xi=\frac{\sigma^2(2^{R_{\rm t}}-1)}{\eta}.
\label{eq:offloading_no_off_power}
\end{align}

We consider two offloading modes. In the first mode, BS~$0$ is allowed to serve $U_1$ using the resource originally assigned to it in Cell~$1$. This can reduce the required transmit power, but the corresponding resource remains occupied in Cell~$1$. In the second mode, $U_1$ is completely offloaded to BS~$0$ and releases its original resource. It then shares a resource in Cell~$0$ with $U_0$, requiring joint pinching-antenna positioning, power allocation, and multiple-access design.

\subsubsection{Traffic-Offloading and Pinching-Antenna Location Design}
We first consider offloading without resource release. Let $P_0$ denote the power used by the pinching antenna to serve $U_0$, while $P_1$ and $P_{01}$ denote the powers used by BS~$1$ and the pinching antenna, respectively, to serve $U_1$. The total transmit-power minimization problem is
\begin{subequations}
\begin{align}
\underset{P_0,P_1,P_{01},\widetilde{x}}{\min}
\quad& P_0+P_1+P_{01} \label{eq:offloading_power_obj}\\
\st\quad& \frac{P_0}{\|\boldsymbol{\psi}_0-\widetilde{\boldsymbol{\psi}}^{\rm Pin}\|^2} \geq \xi, \label{eq:offloading_u0_con}\\
&\frac{P_1}{\|\boldsymbol{\psi}_1-\boldsymbol{\psi}_1^{\rm BS}\|^2}+
\frac{P_{01}}{\|\boldsymbol{\psi}_1-\widetilde{\boldsymbol{\psi}}^{\rm Pin}\|^2}\geq \xi, \label{eq:offloading_u1_con}\\
&
P_0,P_1,P_{01}\geq 0,\quad\widetilde{x}\in\mathcal{X}.\label{eq:offloading_feasible_con}
\end{align}
\end{subequations}
For a given pinching-antenna location, the optimal power for serving $U_0$ is given by
\begin{align}
P_0^\star = \xi\left\|\boldsymbol{\psi}_0-\widetilde{\boldsymbol{\psi}}^{\rm Pin}\right\|^2.
\label{eq:offloading_u0_power}
\end{align}
The minimum power required to serve $U_1$ is given by
\begin{align}
P_1^\star+P_{01}^\star = \xi\min\left\{\left\|\boldsymbol{\psi}_1-\boldsymbol{\psi}_1^{\rm BS}\right\|^2,
\left\|\boldsymbol{\psi}_1 - \widetilde{\boldsymbol{\psi}}^{\rm Pin}\right\|^2\right\}.
\label{eq:offloading_u1_power}
\end{align}
Therefore, $U_1$ is completely served by the pinching antenna when the distance between pinching antenna and user is smaller than the BS~$1$-user distance. Otherwise, retaining the original association is more power-efficient. The pinching-antenna location effectively determines whether $U_1$ lies within the service region of BS~$0$.

When complete offloading is preferred, the pinching-antenna location problem reduces to
\begin{align}
\underset{\widetilde{x}\in\mathcal{X}}
{\min} \quad \|\boldsymbol{\psi}_0-\widetilde{\boldsymbol{\psi}}^{\rm Pin}\|^2
+\|\boldsymbol{\psi}_1-\widetilde{\boldsymbol{\psi}}^{\rm Pin}\|^2.
\label{eq:offloading_pa_problem}
\end{align}
If the midpoint of the users' projections lies within the feasible waveguide segment, the optimal pinching-antenna location is given by
\begin{align}
\widetilde{x}^{\star} = \frac{x_0+x_1}{2}.
\label{eq:offloading_midpoint}
\end{align}
Otherwise, the pinching antenna is placed at the feasible endpoint closest to this midpoint. This result shows that pinching-antenna positioning balances the two service links while extending the effective coverage of BS~$0$ toward Cell~$1$.

The first mode reduces transmit power but does not relieve resource congestion at BS~$1$. To release the resource occupied by $U_1$, the user must be completely offloaded and served using a resource of Cell~$0$. Suppose that $U_0$ and $U_1$ share this resource through non-orthogonal multiple access (NOMA). The pinching antenna superimposes their signals, and the user with the stronger pinching antenna link performs successive interference cancellation.

Denote the users with the stronger and weaker pinching antenna links by $U_{\rm s}$ and $U_{\rm w}$, respectively, and let their locations be $[x_{\rm s},y_{\rm s},0]^{\mathsf T}$ and $[x_{\rm w},y_{\rm w},0]^{\mathsf T}$. Assuming that complete offloading is feasible, the joint power and pinching-antenna location design can be reduced to
\begin{align}
\underset{\widetilde{x}\in\mathcal{X}}
{\min} \quad \xi\left(
2^{R_{\rm t}}\|\boldsymbol{\psi}_{\rm s}-\widetilde{\boldsymbol{\psi}}^{\rm Pin}\|^2
+\|\boldsymbol{\psi}_{\rm w}-\widetilde{\boldsymbol{\psi}}^{\rm Pin}\|^2\right).
\label{eq:offloading_noma_problem}
\end{align}
For a sufficiently long waveguide segment, the optimal longitudinal coordinate is given by
\begin{align}
\widetilde{x}^{\star}=\frac{2^{R_{\rm t}}x_{\rm s}+x_{\rm w}}{2^{R_{\rm t}}+1}.
\label{eq:offloading_noma_location}
\end{align}
The corresponding minimum total transmit power is given by
\begin{align}
P_{\rm NOMA}^{\star}={}&
\xi\frac{2^{R_{\rm t}}}{2^{R_{\rm t}}+1}(x_{\rm s}-x_{\rm w})^2+\xi 2^{R_{\rm t}}(y_{\rm s}-\widetilde{y})^2
\nonumber\\
&+\xi (y_{\rm w}-\widetilde{y})^2+\xi(2^{R_{\rm t}}+1)d_v^2.
\label{eq:offloading_noma_power}
\end{align}
Equations~\eqref{eq:offloading_noma_location} and~\eqref{eq:offloading_noma_power} reveal two different geometric roles. The users' projections onto the waveguide determine the optimal pinching-antenna location, while their perpendicular distances from the waveguide determine the stronger user and the corresponding decoding order. Compared with offloading without resource release, this mode requires more involved coordination but can simultaneously reduce the load and release radio resources in Cell~$1$.

\begin{figure}[!t]
    \centering
    \includegraphics[width=0.82\linewidth]
    {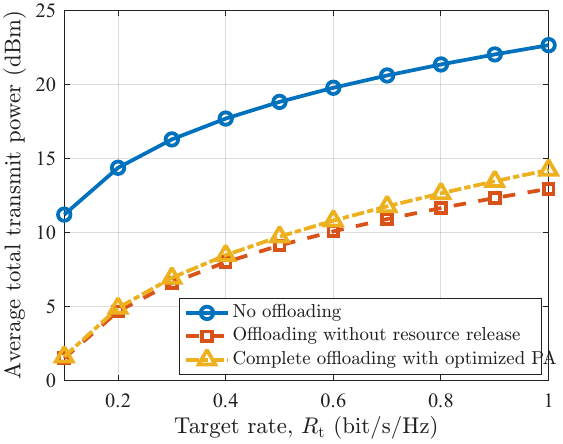}
    \captionsetup{justification=justified,
    singlelinecheck=false,font=small}
    \caption{Average total transmit power versus the target rate for the two-cell traffic-offloading example.} \label{fig:traffic_offloading}
\end{figure}

\subsubsection{Numerical Illustration and Design Insights}
We consider two adjacent cells separated by the boundary $x=0$. BS~$0$ and BS~$1$ are located at $(-40,0,d_v)$ and $(40,0,d_v)$, respectively. BS~$0$ is connected to a $40$-m waveguide extending from its location to the cell boundary, whereas BS~$1$ employs a conventional fixed antenna. User $U_0$ is uniformly generated within $x_0\in[-10,0]$ and $y_0\in[-10,10]$, while the offloading candidate $U_1$ is uniformly generated within $x_1\in[0,10]$ and $y_1\in[-10,10]$. The carrier frequency is $28$~GHz, the antenna height is $d_v=3$~m, and the noise power is $-90$~dBm. The results are averaged over $100$ independent user realizations.

Fig.~\ref{fig:traffic_offloading} compares the average total transmit power of no offloading, offloading without resource release, and complete traffic-and-resource offloading. Both offloading schemes substantially reduce the required power because BS~$0$ can move its pinching antenna toward the cell boundary and establish short links to both users. Their gains over no offloading remain pronounced throughout the considered target-rate range and become particularly evident at high target rates, where serving $U_1$ from the distant BS~$1$ becomes increasingly power-consuming.

Offloading without resource release requires slightly less power than complete traffic-and-resource offloading because $U_1$ can retain its original resource and does not need to share a resource with $U_0$. However, it does not release resources in the overloaded Cell~$1$. Complete traffic-and-resource offloading incurs a moderate additional power cost due to resource sharing and SIC, but it frees the original resource of $U_1$ and therefore provides a more direct means of relieving cell congestion. The preferred mode should consequently be selected according to both transmission power and cell-load requirements.

These results demonstrate that pinching-antenna-assisted traffic offloading can jointly reconfigure serving relationships and effective cell boundaries. More comprehensive analytical and numerical results, including different network geometries, numbers of offloaded users, and resource-release strategies, can be found in~\cite{ding2026impact}.

\subsection{Large-Scale Cell-Boundary Reconfiguration}

The preceding subsections considered a finite number of
neighboring cells, where pinching-antenna positioning modifies desired and
interfering links or changes serving relationships for traffic
offloading. When pinching antennas are deployed throughout a large cellular
network, their collective configuration can have a broader
effect on the spatial organization of the network. In a
conventional deployment, fixed BS locations determine the
Voronoi cells and hence the distributions of cell sizes,
serving distances, and inter-cell interference. Random BS
deployment may therefore produce highly irregular cells with
substantially different boundary distances. Pinching-antenna-enabled BSs
introduce an additional degree of freedom by allowing their
effective radiation locations to be adjusted after the BSs and
waveguides have been deployed. This network-scale effect was
investigated in \cite{zhu2025topological} and is summarized
below using the pinching antenna terminology adopted throughout this
tutorial.

\subsubsection{One-Dimensional Cell-Boundary Analysis}

Consider BSs distributed along a line according to a
homogeneous Poisson point process with density $\lambda_b$.
Under nearest-BS association, the BS locations generate a
one-dimensional Poisson--Voronoi tessellation. Let $L$ denote
the length of a typical cell. Its average value is
\begin{align}
\overline{L}
=
\mathbb{E}[L]
=
\frac{1}{\lambda_b}.
\label{eq:topology_average_cell_length}
\end{align}
Because the seed BS is generally not located at the center of
its cell, the distances from the BS to the two cell boundaries
are unequal. Let $r_{\min}$ and $r_{\max}$ denote the shorter
and longer boundary distances, respectively. Their expected
values are derived in \cite{zhu2025topological} as
\begin{align}
\overline{r}_{\min}
=
\mathbb{E}[r_{\min}]
=
\frac{1}{4\lambda_b},
\qquad
\overline{r}_{\max}
=
\mathbb{E}[r_{\max}]
=
\frac{3}{4\lambda_b}.
\label{eq:topology_boundary_distances}
\end{align}
Thus, the expected distance to the longer boundary is three
times that to the shorter boundary. With a fixed antenna at
the seed BS, this asymmetry produces substantially different
serving distances on the two sides of the cell.

To describe the pinching-antenna  location result without ambiguity, let
$x_e$ denote the coordinate of the pinching antenna measured from the
shorter boundary of the representative cell. For the
interference-free case, the SNR coverage analysis in
\cite{zhu2025topological} shows that the optimal radiation
location coincides with the center of the representative cell,
given by
\begin{align}
x_{e,\mathrm{1D}}^{\star}
=
\frac{
\overline{r}_{\min}
+
\overline{r}_{\max}
}{2}
=
\frac{\overline{L}}{2}.
\label{eq:topology_1d_optimal_location}
\end{align}
Equivalently, since the average coordinate of the original BS
is $\overline{r}_{\min}$, the pinching antenna moves from the BS toward the
longer side by
\begin{align}
\Delta x_{e,\mathrm{1D}}^{\star}
=
x_{e,\mathrm{1D}}^{\star}
-
\overline{r}_{\min}
=
\frac{
\overline{r}_{\max}
-
\overline{r}_{\min}
}{2}
=
\frac{1}{4\lambda_b}.
\label{eq:topology_1d_optimal_displacement}
\end{align}
The physical BS therefore remains fixed, while the pinching antenna moves
the effective radiation location toward the center of the
service region. This compensates for the asymmetry introduced
by random BS deployment and makes the effective cell geometry
more regular.

The analysis in \cite{zhu2025topological} further incorporates
inter-cell interference. In this case, the interference is
separated into that generated by conventional BSs and that
generated by pinching-antenna-enabled BSs, and the corresponding SINR
coverage probability is derived. Moving the pinching antenna changes not
only the desired serving distance but also the interference
relationships with neighboring cells. The analytical and
numerical results show that optimized pinching-antenna positioning
continues to provide rate gains when inter-cell interference
is taken into account.

\subsubsection{Extension to Two-Dimensional Networks}

In a two-dimensional network, randomly distributed BSs
generate irregular polygonal Voronoi cells. Since tractable
distributions of the cell-boundary distances are generally
unavailable, \cite{zhu2025topological} focuses on the
user-association distance and achievable rate. Under
nearest-BS association, the serving distance $r$ has the
probability density
\begin{align}
f_{R_2}(r)
=
2\pi\lambda_b r
\exp\left(-\pi\lambda_b r^2\right),
\qquad r\geq 0,
\label{eq:topology_2d_distance_pdf}
\end{align}
and its average value is
\begin{align}
\overline{r}_2
=
\mathbb{E}[r]
=
\frac{1}{2\sqrt{\lambda_b}}.
\label{eq:topology_2d_average_distance}
\end{align}

Following \cite{zhu2025topological}, let $l_e$ denote the
distance from the BS to the pinching antenna along the waveguide, and let
$d_o$ denote the minimum excitation distance imposed by the
radiation model. For transmit power $P_t$, noise power
$\sigma^2$, and SNR threshold $\gamma_o$, define
$\beta_o=\sigma^2\gamma_o/P_t$. The two-dimensional SNR
coverage probability is
\begin{align}
\mathcal{P}_{\mathrm{SNR}}^{\mathrm{2D}}(\gamma_o)
=
\int_{0}^{\infty}
\exp\left[
-\beta_o
\left(
\max\left\{
|r-l_e|,d_o
\right\}
\right)^2
\right]
f_{R_2}(r)\,
\mathrm{d}r.
\label{eq:topology_2d_snr_coverage}
\end{align}
A closed-form expression for
\eqref{eq:topology_2d_snr_coverage} is derived in
\cite{zhu2025topological}, but directly optimizing this
expression does not yield a simple analytical pinching-antenna location
rule. The random serving distance is therefore replaced by
its mean in \eqref{eq:topology_2d_average_distance}, leading
to the lower-bound-based problem
\begin{align}
l_{e,\mathrm{2D}}^{\dagger}
=
\underset{l_e}{\operatorname{arg\,max}}
\quad
\exp\left[
-\beta_o
\left(
\max\left\{
|\overline{r}_2-l_e|,d_o
\right\}
\right)^2
\right].
\label{eq:topology_2d_approx_problem}
\end{align}
Assuming that the waveguide is sufficiently long to contain
the selected location, the boundary solution adopted in
\cite{zhu2025topological} is
\begin{align}
l_{e,\mathrm{2D}}^{\dagger}
=
\begin{cases}
\overline{r}_2-d_o,
& \overline{r}_2>2d_o,\\
d_o,
& \overline{r}_2\leq 2d_o.
\end{cases}
\label{eq:topology_2d_approx_location}
\end{align}
This solution is obtained from a conservative lower-bound
approximation and is generally slightly smaller than the
numerically optimized pinching-antenna location. The numerical results in
\cite{zhu2025topological} show that it closely approaches the
optimum, particularly for relatively dense BS deployments.

The two-dimensional analysis is also extended to the
interference-limited case by separately characterizing the
interference from conventional BSs and pinching-antenna-enabled BSs.
Although inter-cell interference reduces the average
achievable rate relative to the interference-free case,
optimizing the pinching-antenna locations continues to provide a clear gain
over conventional fixed-antenna deployment. The results also
reveal a direct relationship between BS density and waveguide
length. A denser network has a shorter average serving
distance and therefore requires a shorter waveguide to reach a
favorable pinching-antenna location. Conversely, a sparser network
generally requires a longer waveguide.

\subsubsection{Network-Level Insights}

The main implication is that large-scale pinching-antenna deployment can
modify more than individual propagation links. With fixed
antennas, the irregular geometry produced by random BS
deployment must be accommodated through user association,
scheduling, power control, and interference management.
pinching-antenna-enabled networks can additionally adjust the effective
radiation points associated with this infrastructure, thereby
reducing excessive serving distances and making the effective
cellular geometry more regular. The physical BS topology
remains fixed, while the radio-access topology can be adapted
through pinching-antenna configuration. The analytical and numerical
results in \cite{zhu2025topological} demonstrate that this
capability can improve average achievable rates in both
one- and two-dimensional networks.

Together with the preceding subsections, these results reveal
three progressively broader levels of pinching-antenna-enabled multi-cell
reconfiguration. Pinching-antenna positioning can first reshape desired and
interfering links under fixed serving relationships. It can
then modify serving relationships and cell boundaries for
traffic offloading. When deployed across many cells, it can
further reshape the statistical geometry of the overall
cellular network. As a complementary direction,
stochastic-geometry analysis has also been used to
characterize outage performance in multi-cell pinching-antenna networks
with randomly oriented waveguides, discrete candidate pinching-antenna
locations, inter-cell interference, and blockage
\cite{sun2025stochastic}.

\subsection{Summary and Outlook}
This section has shown that pinching-antenna-enabled multi-cell transmission can introduce spatial reconfigurability at several levels of network operation. Under fixed serving relationships, pinching-antenna positioning can strengthen desired links while reshaping cross-cell interference, thereby reducing the transmit power required for QoS-constrained multi-cell transmission. When serving relationships are also allowed to change, an underutilized cell can move its effective radiation point toward neighboring traffic and support traffic offloading, which makes cell boundaries and resource utilization responsive to spatial load imbalance. At a larger scale, coordinated pinching-antenna deployment can compensate for irregular fixed-site layouts and reshape the statistical geometry of the cellular network. These results show that pinching-antenna reconfigurability extends multi-cell design beyond resource allocation over a fixed radio topology by allowing the effective radiation geometry and serving relationships of the network to be adapted after deployment.

For practical large-scale deployment, pinching-antenna reconfigurability needs to be embedded into the network-control process rather than treated as an isolated optimization variable. This becomes increasingly important when the network is heterogeneous and dynamic, because the preferred radiation locations may change with user mobility, traffic evolution, and the amount of inter-cell information available for coordination. As a result, pinching-antenna configuration should be adapted jointly with the network decisions that determine who is served, where service is provided, and how radio resources are shared across cells.

\section{Region-Level Design for Cell-Free-Inspired
Pinching-Antenna Systems} \label{sec: region level}

Beyond optimizing transmission for a given set of instantaneous users,
pinching-antenna systems can be configured to improve long-term service
across an entire communication region. Their extended waveguides allow
radiation points to be distributed and reconfigured over large spatial
scales, making pinching antennas particularly suitable for region-level
design. The network can configure these radiation points according to
spatially varying service demand and propagation conditions, for
example, to strengthen service in traffic hotspots, improve coverage
balance, avoid severely blocked areas, and enhance service in
weakly covered regions. The design focus therefore shifts from the
performance of individual user links to how wireless service is
distributed across the communication region.

To support such region-wide operation, we consider the
cell-free-inspired pinching-antenna network illustrated in
Fig.~\ref{fig: pa three cases}(c). Multiple waveguides are distributed across
the communication region, connected to a common processing unit, and
coordinated to jointly serve users without relying on rigid cell
boundaries. The cell-free-inspired organization provides centralized
coordination and distributed transmission, while the pinching antennas
add the ability to reconfigure the physical locations from which
signals are radiated. Their combination provides a natural network
architecture for exploiting radiation-location reconfigurability at
the region level.

In the following, we first explain why pinching antennas are
particularly suitable for region-level design under this distributed
architecture. We then consider two representative frameworks, namely
traffic-aware design, which adapts radiation locations to spatial
traffic demand, and geometry-aware design, which accounts for
environmental geometry and blockage conditions.

\subsection{Motivation for Pinching-Antenna-Enabled Region-Level Design}

Conventional cell-free networks distribute a large number of APs across 
a communication region and coordinate them through
centralized processing, thereby providing user-centric service without
rigid cell boundaries \cite{ngo2017cell,bjornson2020scalable}. This
network organization offers several benefits for region-wide service,
including macro-diversity, more uniform coverage, and flexible serving
relationships. A distributed multi-waveguide pinching-antenna system
can retain these benefits because its waveguides can be connected to a
common processing unit and coordinated to jointly serve users across
the region.

The key advantage of pinching antennas for region-level design,
however, lies in their radiation-location reconfigurability. In a
conventional cell-free network, the physical AP locations are generally
fixed after deployment, and network adaptation is mainly achieved
through AP selection, beamforming, power allocation, and user
association. By contrast, an extended waveguide can provide multiple
candidate radiation locations along its deployment path without
requiring a complete AP to be installed at every candidate location.
A pinching-antenna network can therefore adapt not only which
distributed waveguides participate in transmission, but also the
physical locations from which wireless service is provided.

This additional degree of freedom allows the radiation configuration
to be optimized according to spatial information describing the entire
communication region. Rather than repeatedly repositioning the
radiation points for individual instantaneous users, the network can
configure them over a longer time scale according to persistent traffic
patterns and propagation conditions. The resulting configuration can
direct radiation resources toward high-demand areas, improve coverage
balance, or mitigate persistent weak-coverage areas, while
shorter-timescale scheduling, beamforming, power allocation, and other
transmission decisions continue to serve instantaneous users.

The specific form of region-level design depends on the spatial
information available to the network. Traffic-aware design uses
long-term traffic distributions to adapt radiation locations to
spatially nonuniform service demand, whereas geometry-aware design uses
environmental geometry and blockage information to improve region-wide
coverage and robustness. These two representative frameworks are
discussed in the following subsections.

\subsection{Traffic-Aware Region-Level Design}

Wireless traffic is generally distributed nonuniformly across a
communication region. In offices, factories, shopping malls, and other
public spaces, users and service requests may concentrate around
several traffic hotspots, while other areas experience relatively low
demand. Although instantaneous user locations and service requests can
change rapidly, their aggregate spatial distribution often evolves
much more slowly. This difference in time scales provides an
opportunity for traffic-aware pinching-antenna configuration. Optimizing PA
locations according to instantaneous users can strengthen
user-specific links and improve instantaneous transmission
performance, but adapting the pinching-antenna locations to frequent user changes
may require repeated reconfiguration and channel acquisition.
Traffic-aware region-level design complements such instantaneous
optimization by configuring the pinching-antenna locations according to the
long-term spatial distribution of user demand. The resulting radiation
configuration can be maintained over a relatively long time scale,
while scheduling, beamforming, and radio-resource allocation are
optimized based on instantaneous user and channel states. To illustrate
how long-term traffic information can guide pinching-antenna configuration, we next
introduce the system, channel, and traffic models, formulate a
representative pinching-antenna location optimization problem, and present the
corresponding performance results.

\begin{figure}[!t]
    \centering
    \includegraphics[width=0.82\linewidth]
    {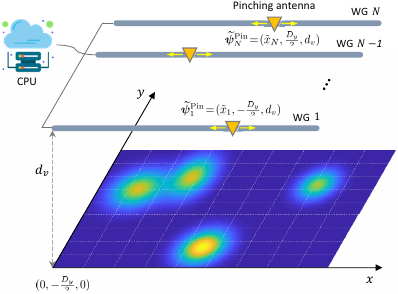}
    \captionsetup{justification=justified,
    singlelinecheck=false,font=small}
    \caption{Illustration of the considered traffic-aware
    pinching-antenna system with multiple traffic hotspots and
    distributed waveguides.}
    \label{fig:traffic_system_model}
\end{figure}

\subsubsection{System, Signal, and Traffic Models}

Consider a rectangular communication region
$\mathcal{R}=[0,D_L]\times[-D_W/2,D_W/2]$, where $D_L$ and
$D_W$ denote its length and width, respectively. As illustrated in
Fig.~\ref{fig:traffic_system_model}, $N$ waveguides are deployed
parallel to the $x$-axis at height $d_v$. The waveguides are uniformly
distributed along the $y$-axis, with the horizontal coordinate of
waveguide $n$ given by
\begin{align}
\widetilde{y}_n
=
-\frac{D_W}{2}+(n-1)d_h,
\ \
d_h=\frac{D_W}{N-1}.
\label{eq:traffic_waveguide_location}
\end{align}
Each waveguide supports one pinching antenna whose location can be continuously
adjusted along the waveguide. The location of pinching antenna $n$ is denoted by
\begin{align}
\widetilde{\boldsymbol{\psi}}_{n}^{\mathrm{Pin}}
=
[\widetilde{x}_n,\widetilde{y}_n,d_v]^{\mathsf T},
\ \
0\leq\widetilde{x}_n\leq D_L,
\label{eq:traffic_pa_location}
\end{align}
and the configurable pinching-antenna locations are collected in
$\widetilde{\mathbf{x}}
=[\widetilde{x}_1,\ldots,\widetilde{x}_N]^{\mathsf T}$. All
waveguides are connected to a common processing unit and jointly serve
users in the communication region.

To describe how user activity is distributed across the communication
region, define $\boldsymbol{U}=[X,Y]^{\mathsf T}$ as the horizontal
location of an active user selected at a scheduling instant. Its spatial
density $f_U(x,y)$ indicates how frequently user activity occurs around
different locations and therefore represents the long-term traffic
pattern. Following \cite{zhang2021predictive}, this density is
approximated by a weighted combination of two-dimensional Gaussian
components as
\begin{align}
f_U(x,y) =\sum_{\ell=1}^{L}\alpha_\ell\mathcal{N}\left([x,y]^{\mathsf T};\boldsymbol{\mu}_\ell,\boldsymbol{\Sigma}_\ell\right),
\label{eq:traffic_map}
\end{align}
where $L$ is the number of traffic hotspots,
$\alpha_\ell\geq0$ is the relative traffic intensity of hotspot
$\ell$, and $\sum_{\ell=1}^{L}\alpha_\ell=1$. Moreover,
$\boldsymbol{\mu}_\ell
=[\mu_{\ell,x},\mu_{\ell,y}]^{\mathsf T}$ specifies the hotspot
center, while $\boldsymbol{\Sigma}_\ell$ characterizes its spatial
spread. This model can represent hotspots with different locations,
traffic intensities, and coverage areas.

For tractable region-level evaluation, the communication region is
divided into a finite set of grids denoted by $\Omega$. Let 
$\boldsymbol{\psi}_{u,v} = [x_u,y_v,0]^{\mathsf T}$ 
denote the center of grid $(u,v)$. Its traffic probability is
approximated as
\begin{align}
p_{u,v} \approx f_U(x_u,y_v)\Delta_x\Delta_y,
\label{eq:traffic_weight}
\end{align}
followed by normalization such that
$\sum_{(u,v)\in\Omega}p_{u,v}=1$, where $\Delta_x$ and
$\Delta_y$ are the grid dimensions. A larger $p_{u,v}$ therefore
indicates that an active user is more likely to appear in grid
$(u,v)$.

The distance between pinching antenna $n$ and grid $(u,v)$ is
\begin{align}
r_{u,v,n}(\widetilde{x}_n)=\sqrt{(x_u-\widetilde{x}_n)^2+(y_v-\widetilde{y}_n)^2+d_v^2}.
\label{eq:traffic_distance}
\end{align}
Let $h_{u,v,n}$ denote the corresponding channel coefficient. Following
the probabilistic LoS model introduced in Sec. \ref{subsec: channel model} and the residual
NLoS model adopted in \cite{xu2026_losnlos}, the channel is represented
as
\begin{align}
h_{u,v,n} = \gamma_{u,v,n}h_{u,v,n}^{\mathrm{LoS}} + h_{u,v,n}^{\mathrm{NLoS}},
\label{eq:traffic_channel}
\end{align}
where $\gamma_{u,v,n}\in\{0,1\}$ indicates whether the LoS path is
available. Its probability is modeled as
\begin{align}
\Pr\{\gamma_{u,v,n}=1\}
=
\exp\left[
-\beta r_{u,v,n}^{2}(\widetilde{x}_n)
\right].
\label{eq:traffic_los_probability}
\end{align}
The LoS channel satisfies
\begin{align}
\left|h_{u,v,n}^{\mathrm{LoS}}\right|^2
=
\frac{\eta}
{r_{u,v,n}^{2}(\widetilde{x}_n)},
\end{align}
while the residual NLoS component is modeled as
\begin{align}
h_{u,v,n}^{\mathrm{NLoS}}
\sim
\mathcal{CN}\left(
0,
\frac{\mu_n^2}
{r_{u,v,n}^{2}(\widetilde{x}_n)}
\right),
\end{align}
where $\frac{\mu_n^2}{r_{u,v,n}^{2}(\widetilde{x}_n)}$ denotes the channel power of the NLoS paths.
Consequently, the average channel power is given by
\begin{align}
\mathbb{E}\!\left[|h_{u,v,n}|^2\right]
=
\frac{
\eta\exp\left[
-\beta r_{u,v,n}^{2}(\widetilde{x}_n)
\right]+\mu_n^2}
{r_{u,v,n}^{2}(\widetilde{x}_n)}.
\label{eq:traffic_average_gain}
\end{align}

At each scheduling instant, one active user is served by all $N$ pinching antennas
using maximum-ratio transmission (MRT). Define
$\mathbf{h}_{u,v}
=[h_{u,v,1},\ldots,h_{u,v,N}]^{\mathsf T}$ as the channel vector
for a user located at grid $(u,v)$. The received signal is given by
\begin{align}
z_{u,v}
=
\sqrt{P_{\mathrm T}}\,
\mathbf{h}_{u,v}^{\mathsf T}\mathbf{w}_{u,v}s
+n_{u,v},
\label{eq:traffic_received_signal}
\end{align}
where $\mathbb{E}[|s|^2]=1$,
$n_{u,v}\sim\mathcal{CN}(0,\sigma^2)$, and the MRT beamformer is 
$\mathbf{w}_{u,v} = \frac{\mathbf{h}_{u,v}^{*}}{\|\mathbf{h}_{u,v}\|}$. 
The resulting instantaneous SNR is given by
\begin{align}
\Gamma_{u,v}(\widetilde{\mathbf{x}}) = \frac{P_{\mathrm T}}{\sigma^2}\|\mathbf{h}_{u,v}\|^2.
\end{align}
Averaging over the random LoS and NLoS channel states gives
\begin{align}
\overline{\Gamma}_{u,v}(\widetilde{\mathbf{x}})
=
\frac{P_{\mathrm T}}{\sigma^2}
\sum_{n=1}^{N}
\frac{
\eta\exp\left[
-\beta r_{u,v,n}^{2}(\widetilde{x}_n)
\right]+\mu_n^2}
{r_{u,v,n}^{2}(\widetilde{x}_n)}.
\label{eq:traffic_grid_snr}
\end{align}
This quantity measures the long-term service quality at a particular
grid but does not account for how frequently that grid is occupied.

To incorporate spatially nonuniform demand, the traffic-weighted
network average SNR is defined as
\begin{align}
\overline{\Gamma}_{\mathrm{net}}
(\widetilde{\mathbf{x}})
=
\sum_{(u,v)\in\Omega}
p_{u,v}
\overline{\Gamma}_{u,v}
(\widetilde{\mathbf{x}}).
\label{eq:traffic_network_snr}
\end{align}
This metric approximates the average SNR experienced by an active user
whose location follows the traffic distribution in
\eqref{eq:traffic_map}. Improving the SNR at a high-traffic grid
therefore contributes more to
$\overline{\Gamma}_{\mathrm{net}}$ than providing the same improvement
at a location where users rarely appear.

\subsubsection{Pinching-Antenna Location Optimization}

A representative traffic-aware region-level design is to maximize the
traffic-weighted network average SNR by optimizing the pinching-antenna locations.
The associated problem is formulated as
\begin{subequations}
 \begin{align}
\underset{\widetilde{\mathbf{x}}}{\max}
\quad&
\overline{\Gamma}_{\mathrm{net}}
(\widetilde{\mathbf{x}})
\label{eq:traffic_opt_obj}\\
\st \quad&
0\leq\widetilde{x}_n\leq D_L,
\quad n=1,\ldots,N.
\label{eq:traffic_opt_con}
\end{align}   
\end{subequations}

Unlike link-level pinching antenna optimization for an instantaneous user, this
problem determines a long-term radiation configuration by jointly
considering all representative user locations and their traffic
probabilities. Although problem~\eqref{eq:traffic_opt_obj} is generally nonconvex,
its objective has a useful separable structure. In particular,
rearranging the summations in
\eqref{eq:traffic_network_snr} gives
\begin{align}
\overline{\Gamma}_{\mathrm{net}}
(\widetilde{\mathbf{x}})
=
\sum_{n=1}^{N}f_n(\widetilde{x}_n),
\label{eq:traffic_separable}
\end{align}
where
\begin{align}
f_n(x)
=
\frac{P_{\mathrm T}}{\sigma^2}
\sum_{(u,v)\in\Omega}
p_{u,v}
\frac{
\eta\exp\left[-\beta r_{u,v,n}^{2}(x)\right]+\mu_n^2}
{r_{u,v,n}^{2}(x)}.
\label{eq:traffic_per_waveguide}
\end{align}
The original problem can therefore be decomposed into $N$ independent
one-dimensional problems,
\begin{align}
\underset{0\leq x\leq D_L}{\operatorname{maximize}}
\quad
f_n(x),
\qquad n=1,\ldots,N.
\label{eq:traffic_scalar_opt}
\end{align}

Each one-dimensional objective is a traffic-weighted combination of
the contributions from different hotspots and may contain multiple
stationary points. The method developed in \cite{xu2026environment1}
exploits the structure of these hotspot-induced components to construct
a small set of candidate pinching-antenna locations. It coarsely brackets stationary
points, refines them through bisection, and selects the candidate
providing the largest objective value. This procedure is applied
independently to all waveguides. Further structural analysis and
algorithmic details can be found in \cite{xu2026environment1}.

\subsubsection{Performance Illustration and Design Insights}

\begin{figure}[!t]
    \centering
    \includegraphics[width=0.82\linewidth]
    {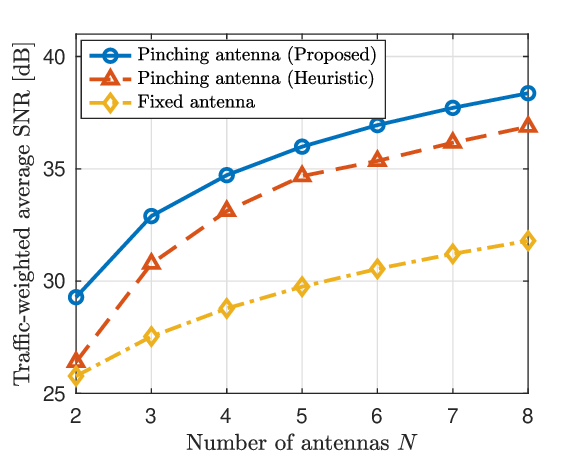}
    \captionsetup{justification=justified,
    singlelinecheck=false,font=small}
    \caption{Traffic-weighted network average SNR versus the number
    of waveguides and pinching antennas.}
    \label{fig:traffic_performance}
\end{figure}

We next use a representative simulation to illustrate how traffic-aware
pinching-antenna positioning improves region-level service. Figure
\ref{fig:traffic_performance} compares the traffic-weighted network
average SNR achieved by the optimized traffic-aware design with two
representative benchmarks. The hotspot-center scheme places the pinching antennas at
the horizontal coordinates of traffic-hotspot centers and represents a
simple traffic-informed heuristic. The fixed-antenna scheme employs an
$N$-element half-wavelength-spaced array centered at $x=D_L/2$, with
its elements distributed along the $y$-axis, and does not adapt its
radiation locations to the traffic map.

The communication region has length $D_L=60$~m and width
$D_W=200$~m, and the waveguides are deployed at height $d_v=10$~m.
The carrier frequency is $28$~GHz, the total transmit power is
$40$~dBm, the noise power is $-70$~dBm, the residual NLoS power is
$-60$~dBm, and the LoS blockage parameter is $\beta=0.01$. The
communication region is divided into $400\times120$ grids. Each
traffic map contains three hotspots whose centers are independently
and uniformly generated within the communication region. Their traffic
intensities are independently generated and then normalized, while all
hotspots share the covariance matrix
\begin{align}
\boldsymbol{\Sigma}
=
\operatorname{diag}
\left(
(0.15D_L)^2,
(0.2D_W)^2
\right).
\end{align}
The results are averaged over $100$ independently generated traffic
maps.

As shown in Fig.~\ref{fig:traffic_performance}, the optimized
traffic-aware design achieves the highest traffic-weighted network
average SNR for all considered numbers of waveguides and pinching antennas. The
hotspot-center scheme also substantially outperforms the fixed-antenna
benchmark, demonstrating the benefit of placing radiation points near
areas with high service demand. However, hotspot-center placement does
not jointly account for hotspot intensity, spatial spread, propagation
distance, and LoS availability. Optimizing the complete
traffic-weighted objective therefore provides a consistent additional
gain.
The traffic-weighted network average SNR increases with the number of
waveguides and pinching antennas for all three schemes because additional radiation
points or antenna elements contribute to the received signal power.
The gain gradually diminishes as $N$ increases. For the pinching antenna schemes,
this indicates that newly added radiation points provide progressively
smaller improvements once the dominant traffic regions are adequately
served. The performance gap between the optimized and hotspot-center
schemes also becomes smaller as more pinching antennas are deployed, since a larger
number of radiation points allows the simple heuristic to cover the
traffic hotspots more effectively. Nevertheless, the optimized design
maintains the best performance throughout the considered range.
The comparison further shows that effective traffic-aware
configuration cannot generally be reduced to aligning pinching antennas with
individual hotspot centers. A high-intensity and concentrated hotspot
may attract a pinching antenna toward its center, whereas several nearby or
overlapping hotspots may be served more effectively from an
intermediate location. The optimized pinching-antenna configuration is therefore
determined by the complete traffic distribution and its interaction
with the propagation environment.

More extensive evaluations can be found in
\cite{xu2026environment1}, including comparisons with exhaustive
position search and projected-gradient optimization, computational
complexity results, the effects of the communication-region size, and
a fairness-oriented design that maximizes the worst average SNR over
regions with non-negligible traffic.

\subsection{Geometry-Aware Region-Level Design}

The spatial traffic distribution indicates where wireless service is
needed, whereas the propagation geometry determines how effectively
that service can be provided. In indoor and blockage-rich
environments, walls, partitions, shelves, and other structures create
location-dependent LoS conditions and persistent weak-coverage areas.
Consequently, two locations with similar service demand may experience
substantially different propagation conditions. Since the physical
layout of such an environment generally changes slowly, its geometry
can be mapped and used to configure the radiation locations over a
relatively long time scale. Geometry-aware region-level design
exploits this information to avoid unfavorable propagation locations,
make better use of available LoS paths, and improve coverage across the
communication region. In this part, we focus on how environmental geometry can
guide pinching-antenna configuration.

\begin{figure}[!t]
    \centering
    \includegraphics[width=0.82\linewidth]
    {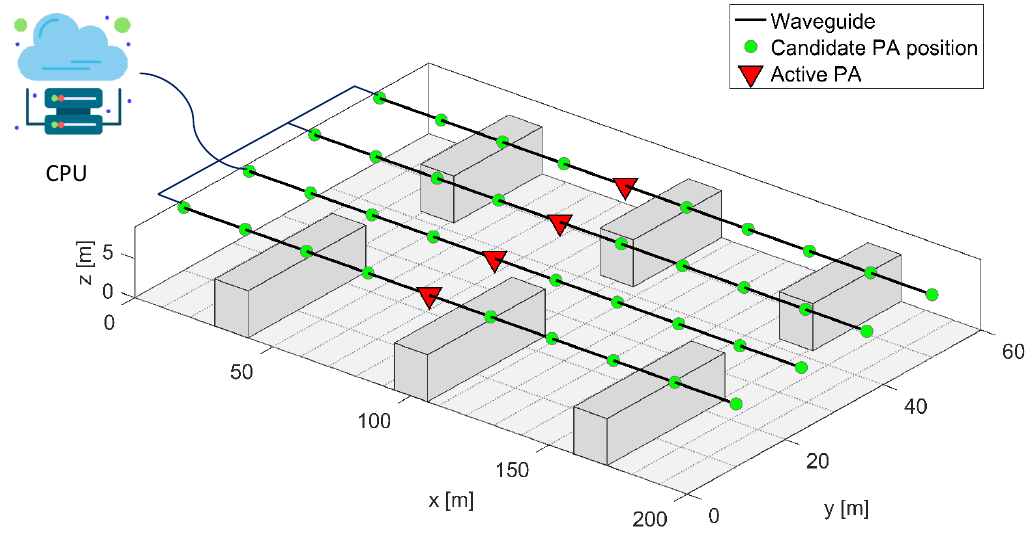}
    \captionsetup{justification=justified,
    singlelinecheck=false,font=small}
    \caption{Illustration of the considered geometry-aware
    pinching-antenna system with explicit obstacles and multiple
    candidate pinching-antenna locations along each waveguide.}
    \label{fig:geometry_system_model}
\end{figure}

\subsubsection{System, Channel, and Signal Models}

Consider the multi-waveguide pinching-antenna system illustrated in
Fig.~\ref{fig:geometry_system_model}. The communication region and
waveguide deployment follow those introduced in the preceding
subsection. In the geometry-aware design, each waveguide has a finite
set of $M$ candidate pinching-antenna locations. The location of candidate $m$ on
waveguide $n$ is denoted by
\begin{align}
\widetilde{\boldsymbol{\psi}}_{n,m}^{\mathrm{Pin}}
=
[\widetilde{x}_{n,m},\widetilde{y}_n,d_v]^{\mathsf T},
\ \
0\leq\widetilde{x}_{n,m}\leq D_L.
\label{eq:geometry_candidate_location}
\end{align}
This discrete model represents a practical implementation in which
mechanical or electrical activation points are predefined along each
waveguide.

Let $a_{n,m}\in\{0,1\}$ denote the activation state of candidate
$(n,m)$, where $a_{n,m}=1$ indicates that the candidate is activated.
Exactly one pinching antenna is activated on each waveguide. As a result, we have
\begin{align}
\sum_{m=1}^{M}a_{n,m}=1,
\ \  n=1,\ldots,N.
\label{eq:geometry_one_pa}
\end{align}
The activation variables are collected in
$\mathbf{A}=[a_{n,m}]\in\{0,1\}^{N\times M}$. Each row of
$\mathbf{A}$ therefore specifies the active radiation location on the
corresponding waveguide.

Similar to the traffic-aware case, the communication region is divided into a set of grids denoted by $\Omega$, where $\boldsymbol{\psi}_{u,v} = [x_u,y_v,0]^{\mathsf T}$
denotes the center of grid $(u,v)$. To describe the propagation
environment, consider $K$ cuboid obstacles. The region occupied by
obstacle $k$ is modeled as
\begin{align}
\mathcal{B}_k
=
[x_k^{\min},x_k^{\max}]
\times
[y_k^{\min},y_k^{\max}]
\times
[0,H_k],
\label{eq:geometry_obstacle}
\end{align}
where $H_k<d_v$. Grids located inside the obstacle footprints are
excluded from $\Omega$ because they do not belong to the service
region.

For every candidate pinching-antenna location and valid grid, the LoS condition is
determined directly from the obstacle geometry. Define
$\gamma_{u,v,n,m}\in\{0,1\}$ as
\begin{align}
\gamma_{u,v,n,m}
=
\begin{cases}
1, & \text{if the path between candidate $(n,m)$ and} \\
   & \text{grid $(u,v)$ does not intersect any obstacle},\\
0, & \text{otherwise}.
\end{cases}
\label{eq:geometry_visibility}
\end{align}
Unlike the probabilistic LoS state considered in the traffic-aware
example, $\gamma_{u,v,n,m}$ is deterministic once the obstacle layout
and candidate pinching-antenna locations are specified. 
Let $h_{u,v,n,m}$ denote the corresponding channel coefficient. The
geometry-aware LoS/NLoS channel is modeled as
\begin{align}
h_{u,v,n,m}
=
\gamma_{u,v,n,m}h_{u,v,n,m}^{\mathrm{LoS}}
+
h_{u,v,n,m}^{\mathrm{NLoS}},
\label{eq:geometry_channel}
\end{align}

For a given activation matrix $\mathbf{A}$, the effective channel from
waveguide $n$ to grid $(u,v)$ is
\begin{align}
\widetilde{h}_{u,v,n}(\mathbf{A})
=
\sum_{m=1}^{M}
a_{n,m}h_{u,v,n,m}.
\label{eq:geometry_effective_channel}
\end{align}
Collecting the effective channels from all waveguides gives
\begin{align}
\widetilde{\mathbf{h}}_{u,v}(\mathbf{A})
=
[
\widetilde{h}_{u,v,1}(\mathbf{A}),
\ldots,
\widetilde{h}_{u,v,N}(\mathbf{A})
]^{\mathsf T}.
\end{align}
At each scheduling instant, one user is served by the activated pinching antennas
using MRT. The received signal at grid $(u,v)$ is
\begin{align}
z_{u,v}
=
\sqrt{P_{\mathrm T}}\,
\widetilde{\mathbf{h}}_{u,v}^{\mathsf T}(\mathbf{A})
\mathbf{w}_{u,v}(\mathbf{A})s
+n_{u,v},
\label{eq:geometry_received_signal}
\end{align}
where $\mathbb{E}[|s|^2]=1$,
$n_{u,v}\sim\mathcal{CN}(0,\sigma^2)$, and 
$\mathbf{w}_{u,v}(\mathbf{A}) = \frac{\widetilde{\mathbf{h}}_{u,v}^{*}(\mathbf{A})
}{\|\widetilde{\mathbf{h}}_{u,v}(\mathbf{A})\|}$ is the MRT beamformer. 
The corresponding instantaneous SNR is given by
\begin{align}
\Gamma_{u,v}(\mathbf{A}) =
\frac{P_{\mathrm T}}{\sigma^2}
\left\|
\widetilde{\mathbf{h}}_{u,v}(\mathbf{A})
\right\|^2.
\end{align}
Averaging over the random NLoS components yields
\begin{align}
\overline{\Gamma}_{u,v}(\mathbf{A}) = \frac{P_{\mathrm T}}{\sigma^2} \sum_{n=1}^{N}\sum_{m=1}^{M}
a_{n,m}\overline{G}_{u,v,n,m},
\label{eq:geometry_grid_snr}
\end{align}
where $\overline{\Gamma}_{u,v}(\mathbf{A})$ denotes the average SNR \cite{xu2026environment2}.
All terms $\overline{G}_{u,v,n,m}$ depend only on the candidate
locations, obstacle geometry, and channel parameters. They can
therefore be calculated offline and stored as a geometry-dependent
SNR map. Once this map is available, the online design only needs to
select one candidate pinching-antenna location on each waveguide.

To evaluate region-wide coverage, let $\Gamma_{\mathrm{th}}$ denote
the required average SNR. Grid $(u,v)$ is considered covered if
$\overline{\Gamma}_{u,v}(\mathbf{A})\geq\Gamma_{\mathrm{th}}$.
Its coverage indicator is accordingly defined as
\begin{align}
c_{u,v}(\mathbf{A})
=
\mathbbm{1}\left\{
\overline{\Gamma}_{u,v}(\mathbf{A})
\geq\Gamma_{\mathrm{th}}
\right\},
\label{eq:geometry_coverage_indicator}
\end{align}
where $\mathbbm{1}\{\cdot\}$ denotes the indicator function. Assuming
that user locations are uniformly distributed over the valid service
region, the coverage probability is given by
\begin{align}
P_{\mathrm{cov}}(\mathbf{A})
=
\frac{1}{|\Omega|}
\sum_{(u,v)\in\Omega}
c_{u,v}(\mathbf{A}).
\label{eq:geometry_coverage_probability}
\end{align}
This metric represents the fraction of the service region whose
average SNR satisfies the prescribed requirement.

\subsubsection{Pinching-Antenna Activation Optimization}

A representative geometry-aware design is to maximize the number of covered
grids by selecting one candidate pinching-antenna location on each waveguide. The
associated problem is formulated as
\begin{subequations}\label{eq:geometry_opt}
\begin{align}
\underset{\mathbf{A}}{\max}
\quad&
\sum_{(u,v)\in\Omega}
c_{u,v}(\mathbf{A})
\label{eq:geometry_opt_obj}\\
\st\quad&
\sum_{m=1}^{M}a_{n,m}=1,
\ n=1,\ldots,N,
\label{eq:geometry_opt_one_pa}\\
&
a_{n,m}\in\{0,1\},
\ n=1,\ldots,N,\ m=1,\ldots,M.
\label{eq:geometry_opt_binary}
\end{align}
\end{subequations}
Unlike the traffic-weighted objective considered previously, this
formulation assigns equal importance to all valid grids and seeks to
maximize the fraction of the service region meeting a prescribed SNR
requirement.

Problem~\eqref{eq:geometry_opt} is challenging because it contains
discontinuous coverage indicators and binary activation variables. As shown in \cite{xu2026environment2}, an
equivalent mixed-integer linear programming formulation can be
obtained by introducing binary coverage variables and linking them to
the per-grid SNR constraints. Such a formulation provides an exact
solution for moderate problem sizes but may become computationally
expensive for a large grid or a dense candidate set.
A lower-complexity coordinate-ascent method was proposed in \cite{xu2026environment2} to solve problem \eqref{eq:geometry_opt} efficiently.

\subsubsection{Performance Illustration and Design Insights}

\begin{figure}[!t]
    \centering
    \includegraphics[width=0.82\linewidth]
    {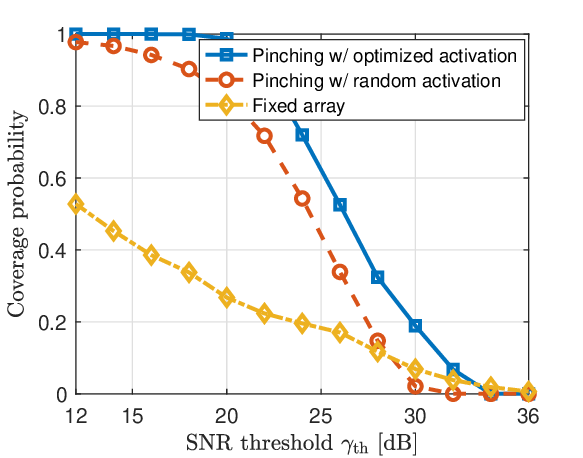}
    \captionsetup{justification=justified,
    singlelinecheck=false,font=small}
    \caption{Average-SNR-threshold coverage probability versus the
    required SNR $\Gamma_{\mathrm{th}}$ under explicit blockage
    geometry, where $N=4$ and $M=10$.}
    \label{fig:geometry_performance}
    \vspace{-3mm}
\end{figure}

We next use a representative simulation to illustrate the coverage
gain provided by geometry-aware pinching-antenna activation. Figure
\ref{fig:geometry_performance} compares three antenna configurations.
The optimized pinching-antenna scheme selects the active radiation locations using
the coordinate-ascent method described above. The random-activation
scheme selects one candidate on each waveguide uniformly at random and
represents pinching-antenna deployment without geometry-aware optimization. The
fixed-antenna scheme employs an $N$-element half-wavelength-spaced
array at the center of the communication region, with its elements
deployed at the same height as the waveguides.

The communication region has length $D_L=200$~m and width
$D_W=60$~m and is divided into $400\times120$ grids. The carrier
frequency is $28$~GHz, the total transmit power is $40$~dBm, the noise
power is $-90$~dBm, and the residual NLoS power is $-60$~dBm. The
waveguides are deployed at height $d_v=10$~m. There are $N=4$
waveguides and $M=10$ candidate pinching-antenna locations on each waveguide. The
environment contains six cuboid obstacles with height $6$~m and width
$8$~m, arranged at different locations within the communication
region. Coverage is evaluated over the valid grids after excluding
those located inside the obstacle footprints.

As shown in Fig.~\ref{fig:geometry_performance}, the coverage
probability decreases as $\Gamma_{\mathrm{th}}$ increases because
fewer grids can satisfy a more stringent SNR requirement. The
geometry-aware design achieves the highest coverage probability over
most of the considered threshold range. By selecting radiation
locations according to the precomputed LoS and distance maps, it can
exploit favorable propagation paths and reduce the number of grids
affected by severe blockage or long propagation distances.
Random pinching-antenna activation generally provides intermediate performance,
showing that distributed and reconfigurable radiation points offer
spatial diversity even without explicit geometry-aware optimization.
However, random activation cannot systematically avoid unfavorable
candidate locations or direct radiation toward poorly covered areas.
The fixed array consequently provides lower coverage at low and
moderate SNR thresholds because its compact and nonreconfigurable
deployment cannot adapt to the spatially varying blockage conditions.
At very high SNR thresholds, the fixed array may provide comparable or
slightly higher coverage. Its antenna elements are concentrated within
a compact aperture at the center of the region and can therefore
produce a strong SNR peak over a small nearby area. In contrast, the
optimized pinching-antenna configuration distributes its radiation points across the
region to increase the total area satisfying a practical target SNR.
Geometry-aware coverage maximization therefore seeks to expand the
covered region at the prescribed threshold rather than maximize the
peak SNR at a few locations.

More extensive evaluations can be found in
\cite{xu2026environment2}, including the effects of increasing the
numbers of waveguides and candidate pinching-antenna locations, the residual NLoS
power, and the transmit power. The reference also considers a
fairness-oriented design that maximizes the worst-grid average SNR and
uses spatial SNR maps to illustrate how different region-level
objectives produce different pinching-antenna activation patterns.

\subsection{Summary and Outlook}

The traffic-aware and geometry-aware designs illustrate two
complementary ways of extending pinching antenna optimization from individual links
to an entire communication region. Traffic-aware design uses the
long-term spatial distribution of service demand to configure
radiation locations and improve the expected service quality
experienced by active users. Geometry-aware design uses explicit
obstacle information to select radiation locations and improve
region-wide coverage under spatially varying propagation conditions.
Although the two designs rely on different spatial information and
performance metrics, they share the same underlying principle of
configuring radiation locations according to region-level information
and service objectives rather than only the instantaneous performance
of individual links.

The representative examples above isolate the benefit of
radiation-location reconfigurability by optimizing pinching-antenna locations or
activation states under fixed waveguide deployments and
radio-resource configurations. Region-level pinching-antenna design can, however,
exploit a much broader set of design variables. At the infrastructure
level, the number, positions, orientations, heights, and lengths of
the waveguides can be optimized according to the service region,
installation constraints, traffic distribution, and environmental
geometry. The segmentation of each waveguide, the number and
distribution of candidate pinching-antenna locations, and the number of
simultaneously activated pinching antennas can also be designed to balance service
performance, implementation cost, and control complexity.

Region-level pinching-antenna configuration can also be jointly optimized with
radio-resource allocation. For example, transmit power can be
allocated across waveguides or active pinching antennas according to their traffic
loads and contributions to region-wide coverage. Pinching-antenna configuration can
further be coordinated with beamforming, bandwidth allocation, user
scheduling, and user association. These decisions naturally operate
over different time scales. Waveguide deployment is generally
determined during network planning, pinching-antenna locations or activation states
can be adjusted according to relatively slow variations in traffic
and environmental conditions, and radio resources can be adapted to
instantaneous users and channel states. Developing scalable
multi-timescale methods that coordinate these decisions is therefore
an important research direction.

Another promising direction is to jointly exploit traffic and
environmental geometry information. A traffic map indicates where
wireless service is needed, whereas a geometry map characterizes how
effectively service can be delivered from different radiation
locations. Their joint use would allow the network to configure its
waveguides, radiation locations, and radio resources according to both
service demand and propagation conditions. Further studies are also
needed to address time-varying traffic and environments, imperfect
spatial information, user mobility, large-scale multi-waveguide
coordination, and practical constraints on pinching-antenna reconfiguration and
control. Together, these directions can broaden region-level pinching-antenna design
from radiation-location optimization into a general framework for
planning, configuring, and operating reconfigurable wireless service
regions.

\section{Pinching-Antenna System Realizations and Network Deployment} \label{sec: realization and deployment}

The preceding sections have focused on the network functionalities and
design opportunities enabled by radiation-location reconfigurability.
Translating these opportunities into practical wireless systems also
requires consideration of how pinching antennas can be physically
realized, organized at the system level, and integrated with existing
network infrastructure. Different system realizations and deployment
modes may provide different levels of radiation-location flexibility,
coordination capability, implementation cost, and suitability for
particular communication environments.
In this section, we first introduce the concept of generalized
pinching-antenna systems and illustrate several representative physical
realizations beyond the dielectric-waveguide-based architecture
considered thus far. We then outline possible network deployment modes
and highlight several implementation and deployment considerations.

\subsection{Generalized Pinching-Antenna Systems}

The dielectric-waveguide-based architecture introduced in
Sec.~\ref{sec: PA fundamental} provides the canonical realization of
pinching antennas. In this architecture, an RF signal propagates along
a dielectric waveguide and is coupled into free space by dielectric
particles or other pinching elements placed at configurable
locations. It directly demonstrates the central capability of creating
radiation points close to users or target regions without installing
a complete transceiver at every candidate location.
Nevertheless, defining pinching-antenna systems solely through this
particular realization may restrict their applicable frequency
ranges, installation environments, and implementation options. For
example, the physical dimensions and propagation characteristics of
dielectric waveguides depend on the operating frequency, while some
environments already contain cables, metallic structures, or
distributed antenna infrastructure that may be used to deliver signals
over extended spatial routes. These considerations motivated the
generalized pinching-antenna framework introduced in
\cite{xu2026generalized}.

From a system perspective, generalized pinching-antenna systems
abstract the underlying principle of pinching antennas from the
dielectric-waveguide-based implementation. They use an extended
structure to distribute signals across space and provide radiation or
reception at configurable locations along that structure. By adjusting
these locations according to user locations, service requirements, or
propagation conditions, the system can reconfigure its effective
wireless access geometry after the supporting infrastructure has been
installed. Different realizations may employ different
signal-propagation media, radiation mechanisms, configuration
granularities, and control methods while retaining this common
system-level capability.
This generalized viewpoint broadens the range of possible
implementations and allows the physical realization to be selected
according to the carrier frequency, service-region geometry,
available infrastructure, reconfiguration requirements, and
implementation cost. In addition to dielectric waveguides,
representative realizations include leaky coaxial cables (LCXs), radio
stripes, and pinching-inspired systems based on other conductive or
guiding structures. We discuss these possibilities next.

\subsubsection{Leaky-Coaxial-Cable-Based Realization}

A LCX, also known as a radiating cable,
contains periodic slots through which part of the signal propagating
inside the cable leaks into free space. LCXs have traditionally been
used to provide reliable coverage in tunnels, mines, railway systems,
corridors, and other confined or elongated environments. In a
conventional LCX, the radiating slots and their operating states are
generally determined when the cable is manufactured and installed. By
incorporating controllable slot-blocking structures, or electronic
switches, the effective
radiation locations can be configured according to user locations and
coverage requirements \cite{nagayama2022proposal,asplund2016leaky}. The resulting architecture provides an 
LCX-based realization of generalized pinching antennas
\cite{wang2025generalized,wang2026leaky1,wang2026leaky2}.

To illustrate its distinctive channel characteristics, consider a
single LCX deployed along the $x$-axis at height $d_v$. Its feed point
is located at
$\boldsymbol{\psi}_0^{\mathrm{LCX}}=[0,0,d_v]^{\mathsf T}$, and one
of its candidate slots is activated at
$\widetilde{\boldsymbol{\psi}}^{\mathrm{slot}}
=[\widetilde{x},0,d_v]^{\mathsf T}$. A user is located at
$\boldsymbol{\psi}^{\mathrm U}=[x,y,0]^{\mathsf T}$. The signal
propagation distance inside the LCX and the wireless distance from
the active slot to the user are, respectively,
\begin{align}
\ell
&=
\left\|
\widetilde{\boldsymbol{\psi}}^{\mathrm{slot}}
-\boldsymbol{\psi}_0^{\mathrm{LCX}}
\right\|
=
|\widetilde{x}|,
\nonumber\\
r
&=
\left\|
\boldsymbol{\psi}^{\mathrm U}
-\widetilde{\boldsymbol{\psi}}^{\mathrm{slot}}
\right\|
=
\sqrt{(x-\widetilde{x})^2+y^2+d_v^2}.
\label{eq:lcx_distances}
\end{align}

The signal first propagates from the feed point to the active slot and
then radiates from the slot toward the user. Under the LoS channel
model considered in \cite{wang2025generalized}, the resulting composite
channel is given by
\begin{align}
h^{\mathrm{LCX}}
=
\frac{\sqrt{\eta}}{r}
10^{-\frac{\kappa\ell}{20}}
\sin(\phi)
\exp\left[
-j\frac{2\pi}{\lambda}
\left(
\sqrt{\varepsilon_r}\ell+r
\right)
\right],
\label{eq:lcx_composite_channel}
\end{align}
where $\kappa$ is the longitudinal attenuation coefficient in
dB/m, $\varepsilon_r$ is the relative permittivity of the cable, and
$\phi$ is the elevation angle between the slot-to-user link and the
horizontal plane. Equation~\eqref{eq:lcx_composite_channel} captures
two consecutive propagation stages. The terms associated with
$\ell$ describe the attenuation and phase accumulation inside the
LCX, whereas those associated with $r$ describe wireless propagation
from the active slot to the user.

A distinctive property of the LCX realization is its directional
slot radiation. For the considered geometry,
$\sin(\phi)=d_v/r$, and hence
\begin{align}
\left|h^{\mathrm{LCX}}\right|^2
=
\eta
10^{-\frac{\kappa\ell}{10}}
\frac{d_v^2}{r^4}.
\label{eq:lcx_channel_power}
\end{align}
Thus, the channel strength depends not only on propagation distance
and in-cable attenuation, but also on the direction from the active
slot to the user. A user located nearly underneath the slot
experiences both a short wireless distance and a large directional
gain, whereas a laterally separated user experiences a longer
distance and a smaller elevation-angle factor.

This directional property can be exploited to support EDMA
\cite{ding2026environment2}. In particular, an active slot can be
selected so that its associated user lies within a favorable
radiation direction, while cross-links toward other users are more
oblique and therefore weaker. LCX-based generalized pinching antennas
can consequently create desired-link and cross-link differences
through the combined effects of radiation-location selection,
distance-dependent path loss, directional leakage, and environmental
blockage. In multiuser operation, slot activation, user assignment,
and power allocation can be further coordinated to exploit these
properties while controlling the remaining interference.

\subsubsection{Radio-Stripe-Based Realization}

A radio stripe (RS) is an elongated cable-like infrastructure
containing multiple active antenna processing units (APUs) connected to a
common processing unit
\cite{interdonato2019ubiquitous,shaik2021mmse}. The shared cable
provides data transfer, synchronization, control signaling, and power
supply, while each APU contains an active radio front end capable of
transmitting, receiving, and locally processing wireless signals.
By selectively activating these APUs, an RS can provide discrete and
electronically controllable radiation or reception points along the
stripe, thereby realizing the principle of generalized pinching
antennas \cite{xu2026generalized-rs}.

To illustrate its channel characteristics, consider an RS deployed
along the $x$-axis at height $d_v$. A candidate APU is located at
$\widetilde{\boldsymbol{\psi}}^{\mathrm A}
=[\widetilde{x},0,d_v]^{\mathsf T}$, while a user is located at
$\boldsymbol{\psi}^{\mathrm U}=[x,y,0]^{\mathsf T}$. Their distance 
is given by
\begin{align}
r = \left\|
\widetilde{\boldsymbol{\psi}}^{\mathrm A}
-\boldsymbol{\psi}^{\mathrm U}
\right\|
=
\sqrt{
(\widetilde{x}-x)^2+y^2+d_v^2
}.
\label{eq:rs_distance}
\end{align}
Since the RS can extend over a large physical aperture and the
activated APU may be located close to the user, the corresponding LoS
channel can be modeled using a spherical wavefront as
\begin{align}
h^{\mathrm{RS}}
=
\frac{\sqrt{\eta}}{r}
\exp\left(
-j\frac{2\pi}{\lambda}r
\right).
\label{eq:rs_channel}
\end{align}
By selecting which candidate APU is activated, the system changes the
APU--user distance and free-space propagation phase and thereby
reconfigures its effective wireless access geometry.

The channel in \eqref{eq:rs_channel} also highlights an important
difference from dielectric-waveguide- and LCX-based realizations. In
an RS, the shared cable does not carry a common RF wave that is
passively coupled into free space. Instead, it distributes data,
control, synchronization, and power to the active APUs. The
APU--user channel therefore does not contain an in-medium RF
attenuation or phase term. Subject to practical synchronization and
calibration, each activated APU can independently control its
transmitted or received signal.

This active realization provides design freedom beyond radiation-point
selection. The network can jointly optimize APU activation, downlink
beamforming, and power allocation or, in the uplink, receive combining
and user power control \cite{xu2026generalized-rs}.
Activating an APU creates an effective
radiation or reception point at its location, whereas deactivating it
removes that point from the current transmission configuration.
However, each active APU also introduces circuit-power, fronthaul,
synchronization, and hardware costs. Activating all available APUs is
therefore not necessarily desirable, which motivates sparse APU
activation jointly designed with radio-resource allocation
\cite{xu2026generalized-rs}.

This viewpoint also distinguishes the generalized pinching-antenna
realization from conventional RS operation. Conventional RS systems
generally treat the deployed APUs as a given distributed antenna
infrastructure and optimize signal processing over that
infrastructure. In the generalized pinching-antenna framework, the
APU activation pattern is itself treated as a system design variable.
The RS can therefore jointly determine where effective wireless access
is provided and how signals are processed at the selected locations.

\subsubsection{Other Pinching-Inspired Realizations}

The generalized principle can also motivate pinching-inspired
realizations that do not rely on dielectric waveguides, LCXs, or
radio stripes. As discussed in \cite{xu2026generalized}, metallic
pipes, wires, engineered surfaces, and other extended conductive
structures may be used to transport signals toward selected service
locations. For example, a signal can propagate along the surface of a
metallic pipe and be radiated into free space through an active antenna
attached or activated near a target user. The active antenna then
plays a role analogous to a pinching point by converting the signal
carried along the structure into a localized wireless link.

These realizations preserve the two-stage interpretation of
pinching-antenna transmission. The first stage transports signals
along an extended physical route, while the second stage establishes
wireless links at selected locations. Depending on the physical
medium, the first stage may rely on guided propagation, surface-wave
propagation, or wired signal distribution, while the second stage may
use passive coupling, directional couplers, electronically switched
radiators, or active antennas. The common objective is to make the
effective wireless access locations configurable without installing
fully independent access points throughout the service region.

Pinching-inspired realizations may be particularly attractive when
existing building, industrial, or transportation infrastructure can
be reused. Their practical performance, however, depends strongly on
the electromagnetic properties of the supporting structure, the loss
of the signal-transport mechanism, the efficiency of signal coupling
and radiation, and the available activation and control hardware.
Their channel and signal models must therefore be developed for the
specific physical platform rather than directly inherited from
dielectric-waveguide-based pinching antennas.

\subsection{Representative Network Deployment Modes}

The preceding subsection describes several physical platforms through
which configurable radiation or reception points can be realized.
How these platforms are incorporated into a wireless network is another important design question. Depending on the service area, available
infrastructure, coverage requirements, and coordination capability,
pinching-antenna systems may operate independently, complement
conventional base stations, or form part of a larger distributed
network. We next discuss three representative deployment modes.

\subsubsection{Standalone deployment} Most existing studies on
pinching-antenna systems have considered a standalone deployment, in
which one or more waveguides are connected to a dedicated BS or
processing unit and directly serve users within a target area
\cite{ding2024flexible,xu2025rate,wang2025antenna,tegos2025minimum,xu2025qos,ouyang2025array,wang2025modeling,tyrovolas2025performance}.
This setting provides a
clear framework for investigating the fundamental benefits of
radiation-location reconfigurability and has been widely adopted for
studying pinching-antenna positioning, beamforming, multiple access, resource
allocation, and other transmission designs. The original
dielectric-waveguide-based implementation also follows this
organization \cite{suzuki2022pinching}. Similar standalone systems can
be realized using LCXs, radio stripes, or other generalized
pinching-antenna structures
\cite{wang2025generalized,xu2026generalized-rs}.

In this deployment mode, the radiation or reception points are
configured according to the users and propagation conditions within
the served area, while signal processing and radio resources are
managed by the associated BS or processing unit. It is suitable for
relatively self-contained environments such as factories, warehouses,
shopping malls, exhibition halls, tunnels, and transportation
facilities. Compared with deploying many independently connected
access points, an extended pinching-antenna structure can distribute
signals and provide multiple candidate wireless access locations
through a common infrastructure. However, its service area and
capacity remain constrained by the extent of the deployed structures
and the available signal-distribution and processing resources.

\subsubsection{Integrated Deployment with Conventional BSs}

Integrating pinching antennas with conventional BSs represents another
important network deployment mode. In this setting, pinching antennas
complement rather than replace conventional BS antennas. Conventional
BSs can continue to provide wide-area coverage, while waveguide-mounted
pinching antennas create additional transmission or reception points in locations
where improved LoS connectivity, coverage extension, traffic
offloading, or spatial cooperation is required. This approach provides
a gradual deployment path because pinching antennas can be introduced in selected
areas without reconstructing the entire cellular infrastructure.

One representative form is cooperative transmission between
conventional BS antennas and waveguide-mounted pinching antennas. In
\cite{zhou2026joint}, the BS and multiple pinching-antenna-equipped waveguides
jointly transmit to users through conventional BS--user and antenna--user
links. Different architectures were considered according to the
locations of the waveguide processing units and the level of
signal-processing cooperation. Separately deployed processing units
provide greater installation flexibility, whereas centralized or
BS-integrated processing enables closer coordination of pinching antenna
positioning, phase alignment, beamforming, and power allocation. The
achievable cooperation gain must therefore be balanced against
synchronization requirements, processing complexity, and the required
modifications to the existing BS infrastructure.

Another possible form uses wireless feeding or relaying to connect a
conventional BS with a remotely deployed pinching antenna structure. The
wireless-fed architecture in \cite{wijewardhana2026wireless} replaces
the fixed wired connection between the BS and the waveguide with a
directional wireless link. This allows the pinching antenna structure to extend
service to locations that cannot be conveniently reached through
physical cabling. A related design in \cite{feng2026hybrid} places a
full-duplex amplify-and-forward relay equipped with directional horn
antennas between the BS and the waveguide input. The BS transmits to
the relay, which amplifies the received signal and feeds it into the
waveguide for radiation through a pinching antenna positioned near the user. This
relay-assisted architecture can improve deployment flexibility and
extend coverage, although its performance is affected by the quality
of the wireless feeding link, relay amplification, residual
self-interference, and additional power consumption. Multi-cell operation provides
a further form of integration with conventional BSs. As discussed in Sec.~\ref{sec: multicell}, pinching antennas can extend
the effective service range of selected BSs to support traffic
offloading or serve as configurable transmission points for
coordinated multipoint transmission
\cite{ding2026impact,amhaz2025comp}. Pinching-antenna configuration can therefore
be coordinated with cell association, load balancing, and inter-cell
cooperation.

\subsubsection{Distributed or Cell-Free-Inspired Deployment}

Pinching antennas can also be organized as a distributed access
infrastructure in which multiple pinching-antenna-equipped waveguides or pinching-antenna-assisted
BSs are deployed across a communication region and coordinated through
a common processing unit. Users can be served by selected waveguides
or jointly by multiple distributed transmission points without
relying on rigid cell boundaries. Unlike the integrated deployment
considered above, this deployment mode uses the distributed pinching antenna
infrastructure itself to provide primary wireless access.

A representative cell-free pinching antenna architecture was investigated in
\cite{li2025cellfree,qian2026pinching,he2026pinching}, where several distributed pinching-antenna-assisted BSs are
connected to a central processing unit and jointly serve multiple
users. Each BS contains multiple waveguides with configurable pinching antennas, and
the digital beamforming and pinching-antenna locations are jointly optimized. The
cell-free organization enables cooperation among geographically
distributed pinching-antenna-assisted BSs, while radiation-location
reconfigurability allows each BS to adapt its effective access
geometry. This architecture therefore combines the macro-diversity of
distributed transmission with the link-level flexibility provided by
configurable radiation locations. As discussed in
Sec.~\ref{sec: region level}, the distributed radiation locations can
also be configured over longer time scales according to region-level
traffic demand, environmental geometry, and coverage requirements
\cite{xu2026environment1,xu2026environment2}.

Different levels of coordination can be adopted according to the
available processing and interconnection capabilities. Fully
centralized processing allows joint pinching-antenna configuration, beamforming,
power allocation, and user association across the distributed
waveguides. When full centralization entails excessive processing or
information-exchange overhead, the waveguides can be organized into
local clusters, with each cluster serving nearby users and exchanging
limited information with the others. User-centric operation can also
select a subset of waveguides and pinching antennas for each user according to
channel quality, traffic demand, or propagation conditions. These
alternatives provide different tradeoffs among cooperation gain,
processing complexity, information exchange, and scalability.

\subsection{Implementation and Deployment Considerations}

The preceding realizations and deployment modes demonstrate several
ways of introducing configurable radiation locations into wireless
networks. Translating these concepts into practical systems requires
careful consideration of the waveguide deployment, guiding
architecture, and radiator design. These hardware choices determine
not only the achievable communication performance, but also the
installation cost, reconfiguration capability, and scalability of the
resulting system.

Waveguide deployment is a fundamental network-planning consideration.
The number, length, position, orientation, and height of the
waveguides determine the spatial range over which radiation locations
can be configured and how well these locations match the service
region \cite{lu2025dual}. Longer waveguides can provide access over extended spatial
routes, but may experience greater in-waveguide attenuation and become
more difficult to install and maintain \cite{xu2026pinching-attenuation}. 
Their deployment is also
constrained by the available walls, ceilings, tunnels, and other
infrastructure surfaces. The waveguide layout should therefore be
designed by jointly considering wireless performance, environment constraints, propagation
loss, installation feasibility, and hardware cost.

Segmented and modular architectures provide a possible means of
supporting larger deployments. A long waveguide can also be divided
into multiple shorter segments with separate feed points, controllers,
or signal-processing resources
\cite{ouyang2026swan}. Such segmentation can shorten
the in-waveguide propagation distance and reduce the resulting
attenuation. It can also isolate the signals carried by different
segments, mitigate unwanted signal leakage and reradiation in uplink
transmission, and allow individual segments to be selected or jointly
processed according to user locations and service requirements.
Moreover, the modular structure can simplify system expansion, local
control, and fault isolation because individual segments can be added,
configured, or maintained separately. These benefits come at the cost
of additional feed points, switches, RF connections, and coordination
overhead \cite{ouyang2026maintainability}. Segmentation therefore 
introduces further design variables,
including segment length, feed placement, segment activation,
RF-chain connection, and inter-segment coordination. 

The design of efficient and controllable pinching antenna radiators represents
another important hardware challenge. Their shape, dimensions,
material, orientation, and coupling with the waveguide jointly
determine the amount of guided power radiated into free space and the
resulting radiation pattern. Recent analytical, full-wave simulation,
and measurement studies have shown that radiator geometry has a
significant effect on radiation efficiency and directionality
\cite{li2026geometry,li2026unlocking}. Further hardware development
should therefore improve radiation efficiency and pattern control
while maintaining sufficient bandwidth, mechanical robustness, and
tolerance to fabrication and placement errors.

Other practical considerations include continuous positioning versus
discrete activation, pinching-antenna reconfiguration speed, synchronization and
calibration, channel acquisition, control signaling, energy
consumption, and long-term hardware reliability. Their relative
importance depends on the adopted physical realization and network
deployment mode, and should be evaluated together with the achievable
communication gain.

\subsection{Summary and Outlook}

This section has discussed how radiation-location reconfigurability can
be realized and incorporated into wireless networks. The generalized
pinching-antenna framework extends the underlying principle from
dielectric waveguides to LCXs, radio stripes, and other
pinching-inspired structures. These realizations differ in their
physical mechanisms and hardware capabilities, but all provide
configurable wireless access locations. Depending on the available
infrastructure and service requirements, they can be deployed as
standalone systems, integrated with conventional BSs, or organized as
distributed or cell-free-inspired access networks.

An important research direction is the joint design of the physical
infrastructure and wireless network operation. Beyond pinching-antenna locations and
activation states, the number, length, layout, segmentation, and feed
configuration of the waveguides can be coordinated with beamforming,
power allocation, user association, and other radio-resource
decisions. Such hardware--communication co-design can balance
performance gains against installation cost, in-medium propagation loss,
and coordination complexity.

Further progress also requires efficient and controllable radiators,
experimentally validated channel and hardware models, scalable
channel acquisition and calibration, and practical reconfiguration
and control mechanisms. System-level prototypes and field trials will
be important for evaluating these factors and supporting the
integration of pinching antennas into practical wireless networks.

\section{Future Directions} \label{sec: future direction}

The preceding sections have demonstrated how radiation-location
reconfigurability can support multiple access, multi-cell transmission,
region-level design, and flexible network deployment. Nevertheless,
many questions remain regarding how this new degree of freedom can be
more broadly incorporated into future wireless systems. In this
section, we discuss two categories of research directions. Consistent
with the network-level focus of this tutorial, we first examine how
pinching antennas may enable new forms of wireless network operation,
including mobility management, programmable radio access, environment
knowledge acquisition, map-assisted optimization, and AI-native
control. We then consider their potential integration with selected
emerging wireless technologies, including low-altitude networks,
space--air--ground integrated networks, and integrated sensing and
communication.

\subsection{Pinching Antennas for Future Wireless Networks}

Radiation-location reconfigurability introduces new possibilities for
organizing and operating wireless networks. In addition to adapting
individual links, the network can dynamically modify its effective
service points, serving relationships, and radio topology according to
user mobility, traffic demand, and propagation conditions. We next
highlight several representative research directions arising from
this capability.

\subsubsection{Mobility-Aware Network Operation}

Mobility management in conventional cellular networks is generally
based on mobile users and geographically fixed BSs. As a user moves
away from its serving BS or crosses a cell boundary, beam tracking,
cell reselection, or handover may be required to maintain reliable
service. Pinching antennas introduce an additional option because the
radiation location can be repositioned or switched among candidate
locations as the user moves. The effective service point may therefore
follow the user over part of its trajectory, maintaining a short
propagation distance or favorable LoS connectivity without immediately
changing the serving BS or waveguide.
Initial studies have investigated pinching-antenna reconfiguration for mobile users.
For example, \cite{amhaz2026mobility} jointly optimized beamforming and
pinching-antenna locations over a user trajectory, while
\cite{zhong2026agentic} considered dynamic pinching antenna sliding and activation
under time-varying channels and explicitly accounted for the latency
of the corresponding AI decisions. These studies demonstrated the
importance of adapting pinching-antenna configurations to user mobility, but mainly
focused on link-level transmission and reconfiguration.

At the network level, pinching-antenna reconfiguration should be coordinated with
mobility-management decisions. The network must determine when a
radiation point should follow a user, when service should be
transferred to another pinching antenna or waveguide, and when an inter-BS handover
is required. Trajectory prediction and environment knowledge can
support proactive pinching-antenna configuration, while robust control is needed to
accommodate prediction errors and unexpected blockage. In multiuser
systems, these decisions must also account for competition among users
sharing the same waveguide or pinching-antenna resources. Future research should
therefore jointly design pinching-antenna configuration, user association,
scheduling, and handover across different time scales while accounting
for service continuity, signaling overhead, reconfiguration delay,
hardware limitations, and mobility robustness.

\subsubsection{Environment Knowledge Acquisition and Map-Assisted
Operation}

The traffic- and geometry-aware designs discussed earlier assume that
the required spatial information is available to the network. A
complementary question is how this information can be efficiently
acquired and updated. Channel knowledge maps provide location-specific
propagation information, while the perception embedding map-enabled 
mobile-network framework, referred to as PEMNet, further
combine channel statistics with spatial-temporal traffic knowledge
\cite{zeng2024tutorial,li2026pemnet}. Pinching-antenna systems can support their
construction by collecting location-tagged measurements associated
with different radiation locations. The configurable pinching antennas can also be
directed toward insufficiently sampled or rapidly changing areas,
providing spatially diverse observations without deploying dedicated
measurement nodes at every location.

The resulting maps can in turn guide pinching-antenna positioning and activation,
user association, scheduling, and radio-resource allocation. This
creates a closed-loop process in which pinching-antenna configuration supports
environment-knowledge acquisition, while the updated knowledge
improves subsequent network operation. Future research should
investigate adaptive measurement-location selection, joint channel
and traffic map updating, uncertainty-aware pinching-antenna configuration, and the
tradeoff between measurement overhead and communication performance.

\subsubsection{Network Control and AI-Native Operation}

Pinching-antenna-enabled networks introduce control variables beyond conventional
scheduling, beamforming, power allocation, and user association. The
network may also need to determine radiation locations, active
waveguides or segments, and the resulting serving relationships. These
decisions operate over different spatial and temporal scales. A
hierarchical control architecture may therefore use local controllers
for relatively fast pinching-antenna activation and resource adaptation, while
higher-level controllers coordinate service regions, inter-waveguide
operation, and longer-term network configurations.

AI-native operation provides a possible approach to managing this
complexity. Initial studies have applied learning methods to specific pinching-antenna design tasks, including pinching-antenna placement, activation, beamforming, and power allocation \cite{xie2025gnn,guo2025gpass,xu2026learning}. Moving beyond such task-specific applications, an AI-native pinching-antenna network can continuously acquire environment and traffic information, predict their evolution, and jointly configure its radiation topology and communication resources
\cite{ding2026environment2,fang2025ai}. Important research
questions include how to represent the network state, coordinate
centralized and distributed intelligence, learn across multiple time
scales, and control inference and signaling overhead. Reliable
operation further requires learned policies that can generalize across
environments and remain robust to imperfect observations, unexpected
events, and hardware constraints.

\subsubsection{Learning-Oriented Network Services}
The reconfigurable radiation locations of pinching antennas can also support network services whose performance depends on more than conventional communication metrics~\cite{lin2026tail, lin2025pinching, asaad2026energy}. Federated learning provides a representative example. In synchronous federated learning, the completion time of each communication round is determined by the slowest participating client. A small number of unfavorable wireless links may therefore dominate the wall-clock training time even when the average communication quality is satisfactory.

An initial study investigated this problem by jointly optimizing the pinching-antenna location and client participation in a synchronous federated learning system with non-IID data~\cite{lin2026tail}. By changing the radiation location, the server can reshape the uplink latency distribution and reduce the probability of severe straggler events. The analysis further showed that client participation should balance statistical importance against tail latency. In particular, when the latency gap between fast and slow clients becomes large, the penalty associated with selecting slow clients is amplified by synchronous sampling. Pinching-antenna reconfiguration provides a physical-layer means of reducing this latency imbalance rather than relying only on biased client selection.

These initial results suggest several directions for extending pinching-antenna-assisted learning beyond the single-server setting. Distributed learning over multiple access points or waveguides would require the radiation configuration to adapt to both the spatial distribution of clients and the learning process. The preferred configuration may also evolve from round to round as client participation, channel conditions, and model convergence change. Another important direction is to investigate asynchronous or deadline-constrained learning, where the network does not always wait for the slowest client and the role of pinching-antenna placement may differ from that in synchronous operation. More broadly, future work can study how communication reconfiguration should respond to task-level objectives such as training time, learning accuracy, and service reliability, thereby extending pinching antennas from communication-oriented optimization toward learning-aware network operation.

\subsubsection{Programmable and Virtualized Radio Access Networks}

The large-scale radiation-location reconfigurability of pinching antennas creates
new opportunities for programmable and virtualized radio access
networks. Beyond virtualizing conventional spectrum, computing, and
signal-processing resources
\cite{gudipati2013softran,foukas2017network}, the network can treat
candidate radiation locations as configurable radio resources. Sets of
waveguides and pinching antennas can then be dynamically grouped to form logical
access points, cells, or service zones according to traffic and
service requirements. The same physical pinching-antenna infrastructure may also
support multiple network slices or operators through different
radiation and radio-resource configurations.

Realizing this concept requires suitable abstractions and
orchestration mechanisms for pinching-antenna resources. The utility of a radiation
location depends on the users, propagation environment, interference,
and neighboring pinching-antenna configurations, while the extent to which pinching antennas can
be independently controlled depends on their physical realization and
feeding architecture. Future research should therefore investigate
how pinching-antenna locations, waveguides, RF chains, spectrum, and processing
resources can be jointly virtualized and allocated while maintaining
performance isolation, scalability, and acceptable reconfiguration
overhead.

\subsection{Integration with Emerging Wireless Systems}
In this subsection, we discuss the potential integration of pinching
antennas with selected emerging wireless technologies, including
space--air--ground integrated networks, integrated sensing and
communication, and wireless networks for the low-altitude economy.

\subsubsection{Space--Air--Ground Integrated Networks}

Space--air--ground integrated networks coordinate satellite, aerial,
and terrestrial systems to provide wide-area coverage and flexible
connectivity across heterogeneous network layers
\cite{liu2018sagin,han2026sagin}. Within this architecture, pinching antennas can provide
configurable terrestrial access or relay points that complement the
broad coverage of satellites and aerial platforms. For example,
satellite capacity can be delivered to a region through a pinching-antenna-assisted
terrestrial structure whose radiation locations are adapted to ground
users, traffic demand, and local blockage conditions.

An initial study considered low Earth orbit (LEO)
satellite-to-terrestrial communication through a pinching-antenna
relay system \cite{jiang2026leo}. In this system, the satellite signal
is forwarded through a dielectric waveguide, while the pinching-antenna locations
are configured to serve terrestrial users affected by blockage. Joint
satellite precoding and pinching-antenna positioning were shown to improve energy
efficiency compared with direct transmission and conventional relay
schemes. This result demonstrates the potential of pinching antennas as
location-flexible interfaces between the space and terrestrial network
layers. Research on pinching-antenna-assisted space--air--ground integration nevertheless
remains at an early stage. Future studies can investigate pinching-antenna-assisted
satellite gateways, multi-satellite and multi-pinching-antenna coordination,
mobility-aware pinching-antenna configuration for time-varying LEO coverage, and
joint optimization of serving-layer selection, traffic routing, user
association, and radio resources. Such designs must account for the
large propagation loss, satellite motion, Doppler shifts, blockage,
handover, and heterogeneous control time scales across the space, air,
and ground layers.

\subsubsection{Integrated Sensing and Communication}

Integrated sensing and communication (ISAC) allows the same wireless
infrastructure and radio resources to support data transmission and
environment sensing. The configurable radiation locations of pinching antennas
provide an additional degree of freedom for shaping both communication
links and sensing geometry. By activating pinching antennas at suitable locations,
the system can establish favorable links to communication users while
illuminating sensing targets from different directions. Distributed
waveguides may further provide spatially separated transmitting and
receiving points, which can improve target visibility and reduce the
effects of blockage.

Initial studies have demonstrated these benefits from several
perspectives. The communication--sensing rate region of pinching-antenna systems was
characterized in \cite{ouyang2026isac}, showing that pinching-antenna location
optimization can enlarge the achievable tradeoff region compared with
fixed antennas. A separated architecture using different waveguides
for signal transmission and echo reception was investigated in
\cite{zhang2025isac}, where pinching-antenna locations were optimized to improve
target illumination while satisfying communication requirements.
Related studies have also examined sensing accuracy and multi-waveguide
pinching-antenna architectures for ISAC \cite{ding2025crlb,li2026isac,mao2025multi}.
Future research should extend these initial link-level designs toward
multiuser, multitarget, and region-level ISAC operation. Important
problems include joint waveguide deployment, pinching-antenna configuration,
waveform and beamforming design, target-dependent activation, and
cooperative sensing among distributed waveguides. Pinching-antenna locations may
also be adapted over time to track moving targets or improve the
sensing quality of insufficiently observed areas.

\subsubsection{Wireless Networks for the Low-Altitude Economy}

The development of the low-altitude economy requires reliable wireless
connectivity for UAVs supporting transportation, delivery, inspection,
emergency response, and other aerial services \cite{jiang2025integrated}. 
Unlike conventional
ground users, UAVs move through three-dimensional airspace with
varying heights and dynamic trajectories. Their channels may therefore
experience rapid changes in propagation distance, blockage, and
interference from multiple visible transmission points. pinching-antenna-equipped
waveguides deployed along building facades, rooftops, towers, and
other elevated infrastructure can provide configurable ground-to-air
radiation locations. In particular, combining horizontal and vertical
waveguides allows the network to adapt its effective service points in
both the horizontal and vertical dimensions according to UAV
positions and flight paths.

Initial studies have considered vertical pinching-antenna deployment for
ultra-low-altitude UAV communications and the joint optimization of
UAV delivery routes and pinching-antenna activation
\cite{lv2026pinching,wang2026vertical}. Building on these initial results,
future research should extend pinching-antenna-assisted aerial links toward
large-scale low-altitude network operation. Important problems include
three-dimensional waveguide deployment, trajectory-aware PA
configuration, aerial-user association and handover, and interference
management among multiple UAVs and terrestrial users. Pinching-antenna location optimization
must also account for trajectory uncertainty, dynamic blockage, and
reconfiguration delay. Coordinating UAV trajectories, PA
configurations, and radio resources across different time scales may
therefore provide an important foundation for reliable and scalable
wireless support of the low-altitude economy.

\section{Conclusions} \label{sec: conclusion}

Future wireless networks are expected to serve increasingly diverse
users, traffic patterns, and propagation environments, creating a
growing need for greater flexibility in how wireless service is
provided across space. Motivated by this need, this tutorial has
examined pinching antennas from a network perspective and discussed
how radiation-location reconfigurability can be incorporated into
wireless network operation. After introducing the system architecture
and channel characteristics, we investigated its implications for
multiple access, multi-cell transmission, traffic offloading, and
region-level design. Representative analytical, optimization, and
numerical examples illustrated how pinching-antenna configuration can strengthen
desired links, reshape interference, modify serving relationships, and
adapt wireless service to traffic demand and environmental geometry.

We further discussed different physical realizations and network
deployment modes for translating these capabilities into practical
systems. Beyond the canonical dielectric-waveguide architecture,
generalized pinching-antenna systems may employ different guiding
structures and radiation mechanisms, while pinching antennas may operate as
standalone access infrastructure, complement conventional BSs, or
form distributed and cell-free-inspired networks. Their continued
development requires advances in waveguide deployment, segmentation,
radiator design, propagation-loss management, channel acquisition,
reconfiguration, synchronization, and network control. Overall,
pinching antennas extend wireless-network design by making radiation
locations configurable after infrastructure deployment. Their
potential therefore lies not only in improving individual links, but
also in enabling networks to adapt their effective access geometry to
users, traffic demand, and propagation environments.

\vspace{1cm}


\smaller[1]
\bibliographystyle{IEEEtran}	
\bibliography{ref-pin}

\end{document}